PAPER • OPEN ACCESS

# First order phase transitions and the thermodynamic limit

PAPER

# First order phase transitions and the thermodynamic limit

Uwe Thiele[1,2,3,7], Tobias Frohoff-Hülsmann[1], Sebastian Engelnkemper[1], Edgar Knobloch[4] and Andrew J Archer[5,6]

1 Institut für Theoretische Physik, Westfälische Wilhelms-Universität Münster, Wilhelm Klemm Str. 9, D-48149 Münster, Germany
2 Center of Nonlinear Science (CeNoS), Westfälische Wilhelms-Universität Münster, Corrensstr. 2, D-48149 Münster, Germany
3 Center for Multiscale Theory and Computation (CMTC), Westfälische Wilhelms-Universität, Corrensstr. 40, D-48149 Münster, Germany
4 Department of Physics, University of California, Berkeley, CA 94720, United States of America
5 Department of Mathematical Sciences, Loughborough University, Loughborough, Leicestershire, LE11 3TU, United Kingdom
6 Interdisciplinary Centre for Mathematical Modelling, Loughborough University, Loughborough, Leicestershire, LE11 3TU, United Kingdom
7 http://www.uwethiele.de

**E-mail:** u.thiele@uni-muenster.de

## Abstract

We consider simple mean field continuum models for first order liquid–liquid demixing and solid–liquid phase transitions and show how the Maxwell construction at phase coexistence emerges on going from finite-size closed systems to the thermodynamic limit. The theories considered are the Cahn–Hilliard model of phase separation, which is also a model for the liquid-gas transition, and the phase field crystal model of the solid–liquid transition. Our results show that states comprising the Maxwell line depend strongly on the mean density with spatially localized structures playing a key role in the approach to the thermodynamic limit.

## 1. Introduction

In this paper we revisit the topic of equilibrium first order phase transitions and elaborate on the origin of the famous Maxwell or ‘equal areas’ construction that applies in the thermodynamic limit (TL), i.e. for infinite systems. In particular, we examine in detail how this construction emerges as the system size becomes larger and larger, thereby gaining additional insight into this construction. According to Ehrenfest’s classification, a first order phase transition is characterized by the appearance of a discontinuity in a first derivative of the free energy with respect to some thermodynamic variable, e.g. for the solid–liquid phase transition, the derivative with respect to pressure becomes discontinuous, implying a jump in the density. For a second order transition all first derivatives are continuous and a discontinuity occurs in second derivatives. More modern classifications define first order transitions as transitions that involve latent heat or as transitions where an order parameter changes discontinuously [1, 2].

At a first order phase transition two different phases (e.g. a gas and a liquid with distinct densities) can coexist and the characteristics of the coexisting states can be calculated by performing the Maxwell construction on the relevant thermodynamic quantity plotted as a function of the order parameter. This can for example be the pressure as a function of the volume or an appropriate free energy specified by the constraints on the system, such as conservation of particle number at fixed volume, or similar. Strictly speaking, the Maxwell construction is only valid in the TL. Likewise, the related discontinuities only arise in this limit (i.e. at infinite system size and particle number) and at equilibrium, after all transients have decayed. The equilibrium states and their transitions are then represented in the form of state diagrams showing various thermodynamic quantities as a function of a relevant control parameter and phase diagrams that show the location of the various stable phases in two- or higher-dimensional parameter space.

However, in a system of finite size or when a finite time horizon is considered, metastable states often play an important role and even unstable states may be crucial, as transient states, for extended time periods. The full set of states and their dependence on the various control parameters is conveniently presented in the form of bifurcation diagrams, well known in the context of dynamical systems and pattern formation theory [3–5]. The place of thermodynamic phase diagrams is taken by 'morphological phase diagrams' or state diagrams and stability diagrams [3, 4, 6]. The notion of a Maxwell point is often used in the context of pattern formation in nonconserved systems [7–11] to indicate equal energy states because of its dynamical significance [12, 13]. In this context this notion applies equally to finite and infinite systems [7, 8] although for finite systems it lacks the thermodynamic relevance as the condition for phase coexistence. In the context of buckling the corresponding concept is the Maxwell load [14].

In this paper we show and discuss how the discontinuities in the TL represented by the Maxwell construction arise from the bifurcation diagrams relating stable, metastable and unstable steady states in finite-size systems. We focus on two systems: (i) phase decomposition of a binary liquid mixture and (ii) the liquid to crystalline solid phase transition. We investigate the transitions that occur in the context of the most basic mean-field continuum models for these two different phase transitions, namely, the Cahn–Hilliard equation [15–17] and the phase field crystal (PFC) model (or conserved Swift–Hohenberg equation) [18–20].

Some aspects related to this question have been considered previously, in particular in relation to the nature of some of the states that can arise in finite-size systems in the two-phase region. References [21–24] describe theory and computer simulation results for atomistic models exhibiting gas-liquid, liquid-hexatic and hexatic-solid phase transitions that indicate how the Maxwell construction develops as the system size increases or the temperature decreases. For example, figures 3–5 of [23] compare Monte-Carlo computer simulation results in a finite three-dimensional domain (see also figures 2 and 3 of [22]) with the results from a capillary drop type model, also in a finite domain, with a mean-field expression for the chemical potential $\mu(\rho)$, where $\rho$ is the average density, and describe their dependence on system size and temperature. These results reveal the emergence of five plateaus in $\mu(\rho)$ with increasing $\rho$, corresponding to states referred to as drop, column, sheet/gap, columnar hole, and spherical hole states. This work is reviewed and further discussed in [25]; see also figure 2 of [21] for molecular dynamics simulation results for liquid-hexatic and hexatic-solid phase transitions in a two-dimensional domain. Note that such computer simulations are able to capture all fluctuation effects, although they are normally unable to determine unstable or metastable states. They cannot, therefore, be employed to determine complete bifurcation diagrams that are the aim of our contribution. Reference [26] presents results similar to the simulations, albeit with fewer states, obtained via mean-field models for liquid–liquid phase decomposition (their figure 2 gives $\mu(\phi_0)$, where $\phi_0$ is the average concentration) and dewetting of a thin liquid film (their figure 3 gives $\lambda(h_0)$, where $h_0$ is the mean film height and $\lambda$ plays a similar role to $\mu$, i.e. that of a Lagrange multiplier) in two-dimensional domains.

A system that can be found in many different states is considered in [27]. This paper investigates the influence of external loading on a nano-slab of nickel, modeling individual atoms via an interatomic potential from [28]. Employing continuation techniques the authors follow minima and maxima of the potential energy landscape as a function of the applied loading and relate the immense multitude of states that result to the apparently random response of the nanostructures to the applied load. Like our work here, this study illustrates the utility of continuation methods in complex systems.

This paper is structured as follows: in section 2 we introduce the Cahn–Hilliard and PFC models as well as the numerical approach used to obtain the bifurcation diagrams that form the main contribution of this work. The subsequent section, section 3.1, summarizes the basic thermodynamics of phase decomposition, followed by a discussion of the decomposition of a binary mixture in one spatial dimension (1D, section 3.2). Section 3.3 extends this discussion to two dimensions (2D), highlighting the influence of the various intermediate planforms. This is followed by a brief summary of the phase diagrams for crystallization in 1D and 2D (section 4.1) with a detailed study of the approach to the TL presented in section 4.2 (1D) and section 4.3 (2D). The final section, section 5, contains a brief conclusion and provides an outlook for further work.

## 2. Model equations and numerical approach

### 2.1. Cahn–Hilliard model

The Cahn–Hilliard model corresponds to the simplest phenomenological macroscopic mean-field continuum model for the dynamics of liquid–liquid demixing and also phase separation in binary alloys. It combines aspects of the Landau theory of phase transitions with linear nonequilibrium thermodynamics and may be derived via Onsager's variational principle, i.e. by minimizing the Rayleighian with respect to the relevant fluxes (see [29] and also the appendix of [30]) or from dynamical density functional theory (DDFT) on truncating a gradient expansion of the free energy and assuming the mobility is constant over the range of densities of interest [31].

The Cahn–Hilliard equation can be written in the form of conserved gradient dynamics for a scalar order parameter field $\phi$ [16, 32]

$$\partial_t \phi = \nabla \cdot \left[ Q(\phi) \nabla \frac{\delta F[\phi]}{\delta \phi} \right], \tag{1}$$

where $Q(\phi)$ is a positive mobility function (not relevant for steady states), and the underlying free energy functional is [17]

$$F[\phi(\mathbf{x}, t)] = \int_V \left[ \frac{\kappa}{2} |\nabla \phi|^2 + f(\phi) \right] \mathrm{d}\mathbf{x}, \tag{2}$$

where $V$ is the domain volume and $\mathrm{d}\mathbf{x}$ is a volume element in $V$. Here, the field $\phi(\mathbf{x}, t)$ corresponds to a local concentration, i.e. a scaled linear combination of local particle number densities. The first term in (2) captures the energetic cost of interfaces ($\kappa \geqslant 0$) and the local free energy density is the double-well potential

$$f(\phi) = \frac{\tilde{a}}{2} \phi^2 + \frac{b}{4} \phi^4, \tag{3}$$

obtained on making a Taylor expansion of the true potential about the critical point [33]. Here $b > 0$ while the parameter $\tilde{a}$ can change sign. We denote the temperature at the critical point as $T_c$ and write $\tilde{a} = a(T - T_c)$, $a > 0$. The critical concentration is $\phi_c = 0$. Note that, for uniform concentration systems, the free energy is just expression (3) multiplied by the volume.

Approximating the mobility $Q$ by a constant $Q_c$, proportional to the diffusion coefficient, one obtains from equation (1) the standard form of the Cahn–Hilliard equation [16]

$$\partial_t \phi = -Q_c \Delta [\kappa \Delta \phi - \partial_\phi f] = -Q_c \Delta [\kappa \Delta \phi - \tilde{a} \phi - b \phi^3]. \tag{4}$$

The term in the square brackets represents (in general, in nonequilibrium) a nonuniform chemical potential that is made up of an interfacial contribution proportional to the Laplacian $\Delta \phi$ and the local term $\partial_\phi f(\phi)$.

To study steady states, i.e. time-independent uniform or nonuniform concentration profiles, we set $\partial_t \phi = 0$ in equation (4) and obtain after two integrations

$$\kappa \Delta \phi - \partial_\phi f(\phi) + \mu = 0, \tag{5}$$

where the integration constant $\mu$ represents a Lagrange multiplier to enforce the constraint that the average concentration $\phi_0 = \frac{1}{V} \int_V \phi(\mathbf{x}) \mathrm{d}\mathbf{x}$ takes a specified value. This constraint reflects the fact that the total number of particles of each of the two species in the binary mixture is separately conserved. This conservation stems from the form of the dynamics in equation (1). In the following $\phi_0$ is used as the relevant control parameter for obtaining stationary solutions. The integration constant after the first integration is set to zero as appropriate for systems with no flow across the boundaries. See, e.g. [34–36] for situations where this condition is not fulfilled. We note in passing that in situations where the total concentration is not controlled, the parameter $\mu$ becomes a relevant control parameter representing an external field or imposed chemical potential. However, this changes the properties of the associated bifurcation diagram [26]. Strictly speaking, $\mu$ is actually a scaled chemical potential difference, since the local concentration $\phi$ is a scaled difference in the local densities, although below we refer to it simply as the chemical potential.

Note that the Cahn–Hilliard model can also be thought of as a simple model for gas-liquid phase separation; in this case the order parameter $\phi$ represents a scaled density change from its critical value.

### 2.2. PFC model

The conserved Swift–Hohenberg equation with cubic nonlinearity, also known as the PFC model [19, 37], provides the simplest phenomenological microscopic mean-field continuum description of the dynamics of the transition between a liquid state and a crystalline state. This local (i.e. partial differential) equation may be derived from a truncated gradient expansion in the DDFT description of an undercooled system undergoing crystallization [19, 38–40]. The governing equation also takes the form of conserved gradient dynamics for a scalar order parameter field $\phi$, as in equation (1), this time with the underlying free energy functional

$$\begin{aligned} F[\phi] &= \int \left[ \frac{\phi}{2} [r + (q^2 + \Delta)^2] \phi + \frac{1}{4} \phi^4 \right] \mathrm{d}\mathbf{x} \\ &= \int \left[ \frac{1}{2} (\Delta \phi)^2 - q^2 |\nabla \phi|^2 + \frac{1}{2} (r + q^4) \phi^2 + \frac{1}{4} \phi^4 \right] \mathrm{d}\mathbf{x} \end{aligned} \tag{6}$$

that has higher order spatial derivatives than the Cahn–Hilliard free energy functional (2). The above two forms of the free energy are related by partial integrations with appropriate boundary conditions. Note that the coefficient of $|\nabla\phi|^2$ is now negative and so favors gradients in the order parameter, while the system is regularized by the strictly positive higher order term that limits the steepness of the variations in $\phi$. The parameter $q$ represents the dominant wave number and $r$ is the undercooling. This model has a critical point at $r_c = 0, \phi_c = 0$, and so is expected to fail before this point is reached [38]: in experiments the freezing transition is first order. The origin of this failure lies in the approximations made in deriving the PFC, but for $r \ll r_c$ the PFC model provides a qualitatively correct description of the freezing transition region. The resulting kinetic equation is of sixth order

$$\partial_t \phi = Q_c \Delta[r\phi + (q^2 + \Delta)^2\phi + \phi^3], \tag{7}$$

and we refer to this conserved Swift–Hohenberg equation as the PFC model. Here $Q_c$ again represents a constant mobility. Other sign conventions as well as other nonlinearities are discussed, e.g. in [19, 20, 41, 42]. As before, steady states of (7) are studied by setting $\partial_t \phi = 0$. After two integrations we obtain

$$r\phi + (q^2 + \Delta)^2\phi + \phi^3 = \mu, \tag{8}$$

where the integration constant $\mu$ again represents a Lagrange multiplier for particle number conservation, i.e. the chemical potential. Thus, the steady states of the PFC equation correspond to steady states of a nonconserved Swift–Hohenberg equation. However, the chemical potential $\mu$ plays an important role through the properties of the associated bifurcation diagram [20, 26] unless $\mu$ is set to zero as appropriate for studies of the nonconserved system [8, 43].

### 2.3. Numerical approach

For both of the models studied here, the Cahn–Hilliard and the PFC equations, we are only interested in the various possible steady states and their bifurcations. Branches of steady state solutions are determined using pseudo-arclength path continuation techniques [26, 44, 45] employing the packages AUTO07 [46, 47] for the Cahn–Hilliard equation in 1D and PDE2PATH [48, 49] for the Cahn–Hilliard equation in 2D. The latter is used for the PFC model in both 1D and 2D. In the 2D case, we choose numerical domains $\mathcal{D}^{\mathrm{num}}$, on which we apply the continuation method, that are fractions of the physical domain $\mathcal{D}$ shown in the figures. This fraction is determined by the symmetry of the state considered. This procedure lowers the computational cost, but the results of the accompanying numerical linear stability analysis have to be interpreted with care as only perturbations fulfilling the assumed symmetries are admitted. As a result bifurcations that break the symmetry of the state are not detected.

In particular, in section 3 we employ the Cahn–Hilliard equation to study liquid–liquid phase decomposition in 1D and 2D domains with periodic boundary conditions. Consequently the critical domain size is defined as $L_c = 2\pi/q_+$ with $q_+$ being the upper limiting wave number of the band of wave numbers with positive growth rate in the linear regime. For all calculations, equation (5) is rescaled such that $q_+ = 1$, i.e. $L_c = 2\pi$. In section 3.3 we consider states on a square domain with periodic boundary conditions. For large domains we use an adaptive mesh to guarantee the convergence of the results and improve computational performance. Since the drop and stripe states under consideration fulfill reflection symmetry in both directions, the calculation is done on one quarter of the physical domain imposing no-flux boundary conditions. Furthermore, in the case of stripe states we use the translation invariance of these states in one direction and directly calculate the associated branches in the 1D system. However, this procedure fails to capture the linear stability of the stripe state with respect to transverse perturbations. Instead, we use the branches of modulated stripes that connect translation-invariant stripe and drop states obtained in 2D calculations to deduce the full linear stability properties of the translation-invariant stripes.

In section 4 we employ the PFC equation to consider crystallization in 1D and 2D. Here, the critical domain size is defined as $L_c = 2\pi/q$ with $q$ representing the dominant wave number, where the maximal growth rate first crosses zero, i.e. at the instability onset. The model is scaled so that $q = 1$ and therefore once again $L_c = 2\pi$. In section 4.3 we analyze localized and periodic states with hexagonal symmetry (see, e.g. figure 12 below). This allows us to use a numerical domain $\mathcal{D}^{\mathrm{num}}$ defined as a 1/12 angular section (angles from 0 to $\pi/6$) of $\mathcal{D}$ with no-flux boundary conditions. This implies that we impose a $\pi/3$ rotational symmetry on the states considered as well as reflection symmetry with respect to the median lines. At small amplitude the periodic hexagonal states can be thought of as resulting from the superposition of three harmonic modes, all with the same wave number $q = 1$ but orientations rotated by $2\pi/3$. In this case it is more convenient to characterize the domain size by the length of the hexagon side. The critical side length for pattern formation is $L_c^{\mathrm{h}} = L_c/\cos(\pi/6) = 4\pi/\sqrt{3}$, i.e. $n$ peaks fit along one side of a hexagonal domain with side length $L_{\mathrm{h}} = nL_c^{\mathrm{h}}$. In section 4.3 we also consider hexagonal front states on a rectangular domain with periodic boundary conditions. A hexagonal structure on a rectangular domain is composed of rhombi invariant under translations by $2L_c$ in the $x$ direction and by $(2/\sqrt{3})L_c$ in the $y$ direction. To discuss hexagonal front states we choose $\mathcal{D}^{\rm num} = L_x^{\rm num} \times L_y^{\rm num}$ with $L_x^{\rm num} = N \cdot L_x^{\rm c} = N \cdot 2L_{\rm c} = 4\pi N$ for different $N$ and fix $L_y^{\rm num} = L_y^{\rm c} = (2/\sqrt{3})L_{\rm c} = 4\pi/\sqrt{3}$.

Since the Cahn–Hilliard and PFC models are both continuity equations and we only consider steady states with periodic or no-flux boundary conditions, we can use the integrated equations (5) and (8), respectively. There are then two possibilities for studying the dependence of the various steady states on the mean concentration $\phi_0$ [26]. Direct continuation in the control parameter $\phi_0$ is incorporated through an integral condition. This additional equation forces the use of an additional free parameter (the Lagrange multiplier $\mu$) which is adapted as part of the continuation procedure. Alternatively, the parameter $\mu$ is employed directly as a control parameter without the need to include any further condition on the continuation procedure. However, an integral has to be evaluated at each $\mu$ in order to determine the corresponding solution measure $\phi_0$. The two approaches lead to the same set of steady states, although their arrangement into solutions branches depends on which parameter is used as the control parameter. This reflects the different stability properties obtained through the different procedures (see conclusion of [20]). Here, only the first approach results in the correct stability properties corresponding to the original conserved dynamics described by equations (4) and (7), respectively. In contrast, the second approach allows for perturbations which alter the mean concentration at fixed imposed chemical potential (e.g. via an external reservoir), a situation that does not correspond to conserved dynamics.

## 3. Liquid–liquid phase decomposition

### 3.1. Phase behavior

First, we review the phase behavior of a binary mixture as described by the free energy (2). We minimize $F[\phi]$ under the constraint of fixed total number of particles, i.e. at fixed mean concentration $\phi_0$. This is equivalent to requiring the grand potential

$$\Omega[\phi] = F[\phi] - \mu \int \phi(\mathbf{x})\mathrm{d}\mathbf{x} \equiv \int \omega \mathrm{d}\mathbf{x} \tag{9}$$

to be minimal. We consider passing to the TL, i.e. the limit of taking the system volume $V = L^d$ to infinity, where $d$ is the dimension of the system, whilst proportionally increasing the particle number, so that $\phi_0$ remains constant as the limit is taken. In the TL the contribution to the free energy due to any interfaces can be neglected, since their contribution scales as $L^{d-1}$, and so the condition for a minimum becomes

$$\partial_\phi f(T, \phi) - \mu \equiv a(T - T_{\rm c})\phi + b\phi^3 - \mu = 0. \tag{10}$$

This condition relates the Lagrange multiplier $\mu$ to $\phi$.

For $T > T_{\rm c}$, equation (10) has only one solution, i.e. the free energy is minimized by the homogeneous state $\phi = \phi_0 \equiv \phi_{\rm h}$, from now on indicated by the subscript h. For a given $\phi_{\rm h}$, the chemical potential depends linearly on $T$: $\mu_{\rm h}(T, \phi) = a(T - T_{\rm c})\phi_{\rm h} + b\phi_{\rm h}^3$ and therefore $\partial\mu_{\rm h}/\partial T = a\phi_{\rm h}$. The corresponding free energy per unit volume is $f_{\rm h}(T, \phi) = \frac{1}{2}a(T - T_{\rm c})\phi_{\rm h}^2 + \frac{b}{4}\phi_{\rm h}^4$, and the pressure is $p_{\rm h}(T, \phi) = -\omega_{\rm h}(T, \phi) \equiv -f_{\rm h} + \mu_{\rm h}\phi_{\rm h} = \frac{1}{2}a(T - T_{\rm c})\phi_{\rm h}^2 + \frac{3}{4}b\phi_{\rm h}^4$, where $\omega_{\rm h}$ is the grand potential density. Moreover, the entropy density at fixed concentration is $s_{\rm h} = -\partial f_{\rm h}/\partial T = -\frac{a}{2}\phi_{\rm h}^2$, while the specific heat at fixed concentration is $-T\partial^2 f_{\rm h}/\partial T^2 = 0$. These results are illustrated in figures 1 and 2 using thick solid blue lines.

For temperatures $T$ below the critical temperature $T_{\rm c}$, equation (10) can have three solutions and depending on $T$ and $\phi_0$ the system may phase-separate into regions with concentrations $\phi = \phi_+$ and $\phi = \phi_-$. These coexisting concentration values are called the binodals and are given by a Maxwell (or double-tangent) construction on $f$. This construction results directly from the minimization of $F[\phi]$ at fixed $\phi_0$ and volume $V$ and implies that the chemical potentials $\mu = \partial_\phi f$ and pressures $p = -f + \mu\phi$ in the two phases $\phi_+$ and $\phi_-$ are equal, i.e. that the conditions for thermodynamic equilibrium hold. For the present symmetric potential (3), this results in the binodals

$$\phi_\pm(T) \equiv \pm\phi_{\rm b} = \pm\sqrt{\frac{a(T_{\rm c} - T)}{b}}. \tag{11}$$

These meet one another and the spinodals at the critical point where $\partial_\phi f = \partial_{\phi\phi} f = \partial_{\phi\phi\phi} f = 0$. From now on the values of quantities at the binodals are indicated by the subscript b. Thus at coexistence $\mu_{\rm b} = 0$ and $f_{\rm b} = \omega_{\rm b} = -p_{\rm b} = -a^2(T - T_{\rm c})^2/4b$. These results are illustrated in figures 1 and 2, using thick black lines.

Note that the homogeneous state also exists in the concentration range between the binodals where it is not the thermodynamically (or globally) stable state. When crossing the binodal, the uniform state initially remains metastable, i.e. it represents a local minimum of the free energy, before becoming unstable on crossing the locus of $\partial_{\phi\phi} f = 0$. The corresponding $\phi$ values are called the spinodals. For the present $f$ they are $\pm\phi_{\rm s} = \pm\sqrt{a(T_{\rm c} - T)/3b}$, i.e. at a given $\phi_0$ one finds the temperature at the spinodal $T_{\rm s} = T_{\rm c} - \frac{3b\phi_0^2}{a} \leqslant T_{\rm c}$. These metastable and unstable homogeneous states are illustrated in figures 1 and 2 using blue dashed and dotted lines, respectively, and the spinodal points are marked by filled circles.

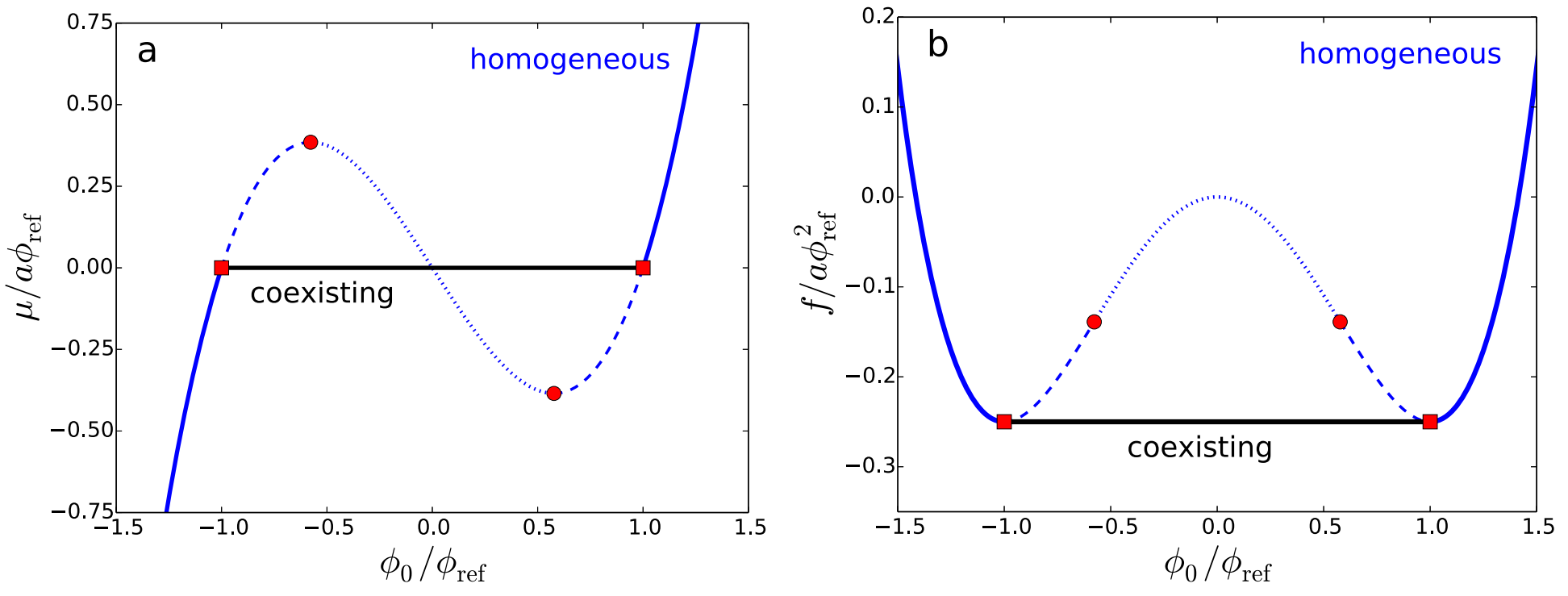

**Figure 1.** (a) The chemical potential $\mu$ and (b) the free energy density $f$ of the homogeneous state (blue lines) and the phase-separated state (thick black line) as a function of (scaled) $\phi_0$, both in the thermodynamic limit at fixed $T$. The values at the binodals and spinodals are indicated by the filled square and circle symbols, respectively. Linearly unstable and metastable homogeneous states are indicated by dotted and dashed lines, respectively. Here $\phi_{\rm ref} = \phi_{\rm b} \equiv \sqrt{a(T_{\rm c} - T)/b}$.

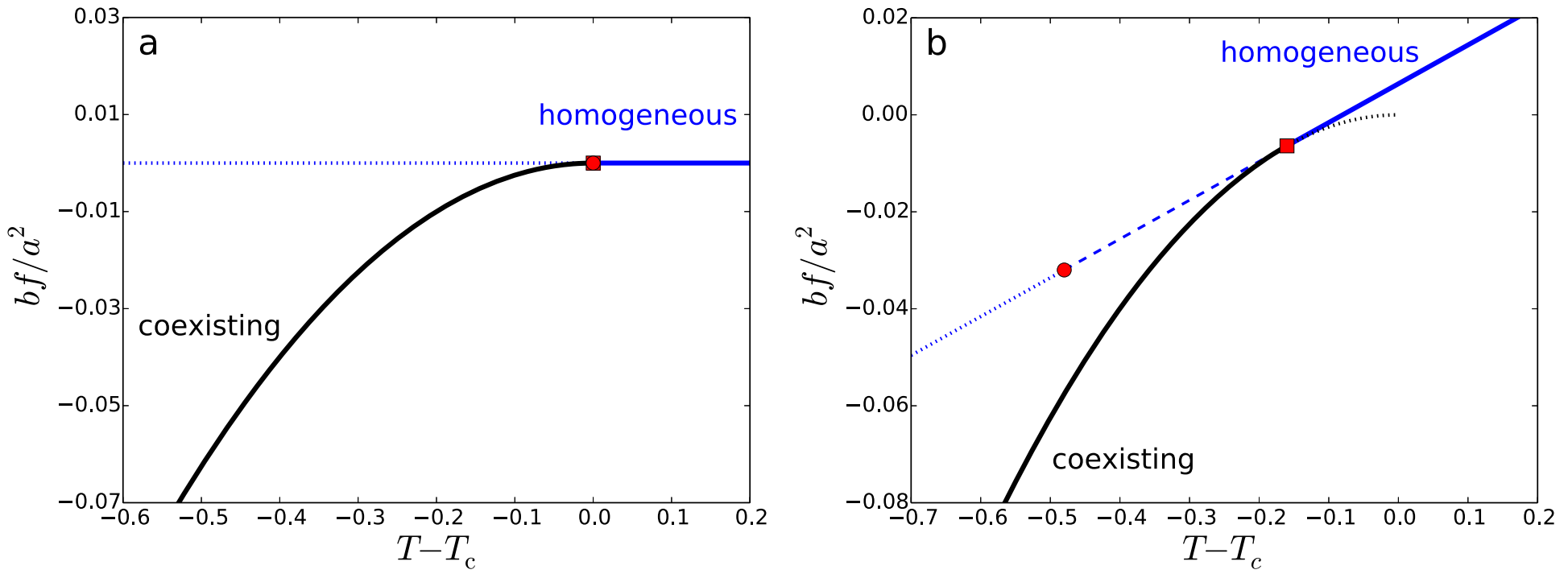

**Figure 2.** The free energy density $f$ of the homogeneous state (blue lines) and of the phase-separated state (thick black line) as a function of the temperature $T$ for (a) the critical density $\phi_0 = \phi_{\rm c} = 0$ and (b) an off-critical density ($\phi_0 = 0.4$). Note that $f(T) = \omega(T)$ for $\phi_0 = 0$ or $\mu = 0$. The values at the binodal ($T_{\rm b}$) and spinodal ($T_{\rm s}$) are indicated by filled square and circle symbols, respectively. Linearly unstable and metastable homogeneous states are indicated by dotted and dashed blue lines, respectively. The part of the binodal that is not accessible for the chosen $\phi_0$ is indicated by a black dotted line. As $\phi_0 \to 0$, both $T_{\rm s}$, $T_{\rm b} \to T_{\rm c}$ and the blue line becomes horizontal.

Owing to the overall concentration constraint, coexisting states can only exist for $\phi_- \leqslant \phi_0 \leqslant \phi_+$. So, for any given $\phi_0$, phase coexistence is only possible below

$$T_{\rm b} = T_{\rm c} - \frac{b}{a}\phi_0^2 \leqslant T_{\rm c}, \tag{12}$$

i.e. when the line $\phi(T) = \phi_0$ crosses the binodal $\phi_{\rm b}(T)$. This condition defines the phase transition temperature $T_{\rm b}(\phi)$ and implies (i) that in the 'critical' case ($\phi_0 = 0$), $T_{\rm s}$, $T_{\rm b}$ and $T_{\rm c}$ all coincide and moreover that the order parameter $\delta\phi \equiv (\phi_{\rm max} - \phi_{\rm min})/2 = \phi_{\rm b} = \sqrt{\frac{a(T_{\rm c} - T)}{b}}$ changes continuously across the (second order) phase transition, while (ii) $T_{\rm s} < T_{\rm b} < T_{\rm c}$ in the 'off-critical' case ($\phi_0 \neq 0$). Then, at $T_{\rm b}$ the order parameter $\delta\phi$ jumps by $\sqrt{\frac{a(T_{\rm c} - T_{\rm b})}{b}} = |\phi_0|$ (first order phase transition). The resulting representation of the phase transition in the phase plane spanned by $\phi_0$ and $T$ and the respective continuous and discontinuous dependence of $\delta\phi$ on $T$ are shown in figure 3. Note that the jump in order parameter is not accompanied by a jump in $\partial f/\partial T$ (see figure 2(b)). This indicates that liquid–liquid phase decomposition is an example where the classical Ehrenfest definition of first order phase transition fails, while that based on a discontinuity in the order parameter holds. The corresponding entropy is $s_{\rm b} \equiv -\partial f_{\rm b}/\partial T = a^2(T - T_{\rm c})/2b = -a\phi_{\rm b}^2/2$, while the specific heat is $-T\partial^2 f_{\rm b}/\partial T^2 = a^2 T/2b$. Finally, $\partial\mu_{\rm b}/\partial T = 0$. Note that this implies $s_{\rm h} > s_{\rm b}$ as $\phi_{\rm h}^2 < \phi_{\rm b}^2$. This is as expected, as in demixing the system reduces its internal energy, paying a trade-off in entropy. As the system is not isolated, this process is accompanied by a flux of heat through the boundary.

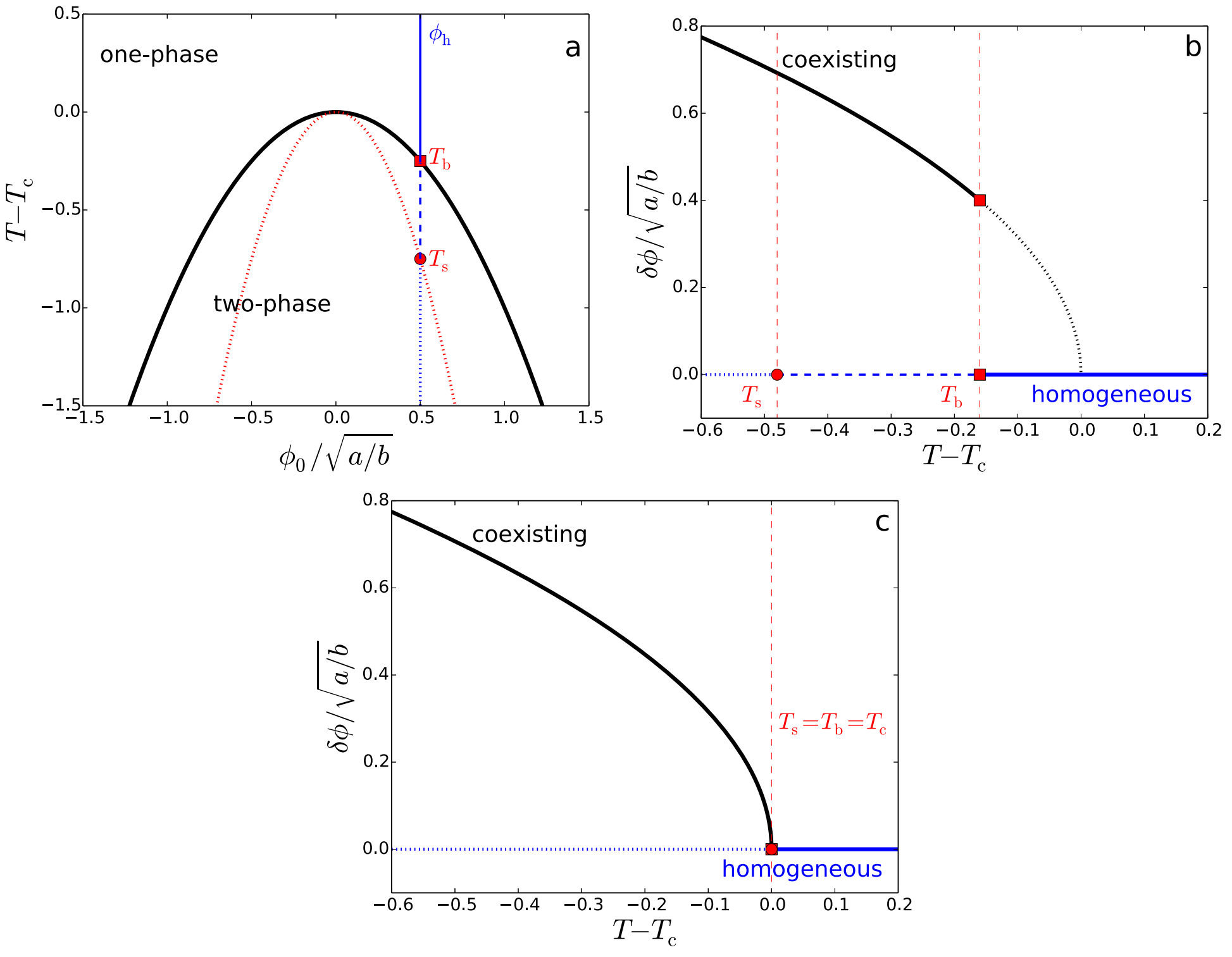

**Figure 3.** Panel (a) shows the liquid–liquid demixing phase diagram in the plane spanned by the mean concentration $\phi_0$ and the temperature $T$. The binodals and spinodals are shown using thick black and thin dashed red lines, respectively. The one- and two-phase regions lie above and below the binodal, respectively. The vertical solid, dashed and dotted lines indicate the range of thermodynamically stable, metastable and unstable homogeneous states for a specific choice of $\phi_h = \phi_0$. The limiting temperatures for this value of $\phi_h$ are indicated by red symbols. Panels (b) and (c) show the dependence of the order parameter $\delta\phi$ on $T$ in the case of a discontinuous transition ($\phi_0 = 0.4$, first order) and a continuous transition ($\phi_0 = 0$, second order), respectively, in the thermodynamic limit. In (b) the states on the binodal that cannot be realized for the chosen $\phi_0$ are indicated by a black dotted line.

### 3.2. Bifurcations in finite domains—phase separation in 1D

In the previous section we described in a compact manner the well-known thermodynamic behavior of a simple binary mixture close to the critical point for liquid–liquid phase decomposition. These results are valid in the TL, i.e. in the limit of diverging system size and particle number where the mean concentration is the relevant control parameter, in addition to the temperature. Implicitly it is also assumed that fluctuations have eliminated all transients, including metastable states. However, real systems have a finite domain size and finite observation times, and in such systems finite-size effects and transients may become important.

Finite systems are often discussed in terms of bifurcation diagrams instead of phase diagrams, following dynamical systems theory and the theory of pattern formation [4, 5]. Here we discuss the behavior of the Cahn–Hilliard equation for a finite domain size combining linear stability and bifurcation analyses, and focus on how the Maxwell construction emerges as the domain size is increased. For simplicity, we present first the case of a 1D domain before moving on to 2D.

In section 3.1 it is mentioned that a homogeneous state $\phi = \phi_h$ loses stability when $\partial_{\phi\phi} f = 0$. The linear dynamic behavior may be studied by introducing the ansatz $\phi = \phi_h + \varepsilon \exp(\lambda t + \mathrm{i}\,\mathbf{q}\cdot\mathbf{x})$ into the kinetic equation (1). Linearizing in $\varepsilon$ gives the dispersion relation

$$\lambda(q) = Q_c \kappa q^2 (q_+^2 - q^2), \tag{13}$$

i.e. the growth or decay rate of a harmonic perturbation as function of its wave number $q = |\mathbf{q}|$. Here $Q_c \equiv Q_c(\phi_h)$ and $\kappa$ are always positive. In contrast, $q_+^2 \equiv -\partial_{\phi\phi} f/\kappa = -[a(T - T_c) + 3b\phi_h^2]/\kappa$ can be positive or negative depending on the curvature of the local free energy density. Note that $\lambda(q)$ is always real since equation (1) has gradient form. The system first becomes unstable when $q_+^2 = 0^+$, which occurs for $\partial_{\phi\phi} f \to 0^-$, corresponding to a long-wave instability. Above the instability threshold, there exists a band of unstable wave numbers $0 < q < q_+$ with the largest $\lambda$ at $q_{\max} = q_+/\sqrt{2}$. The zero crossing at $q = q_+$ corresponds to a steady state bifurcation from the homogeneous state $\phi = \phi_h$.

Fixing the domain size to $L$ selects the wave number $q_L^n = 2\pi n/L$, where $n = 1, 2, \ldots$, and the zero crossing of (13) determines the threshold value of $\phi_L^n$ (at fixed $T$) or of $T_L^n$ (at fixed $\phi$). Here, we focus on the first option and obtain the thresholds

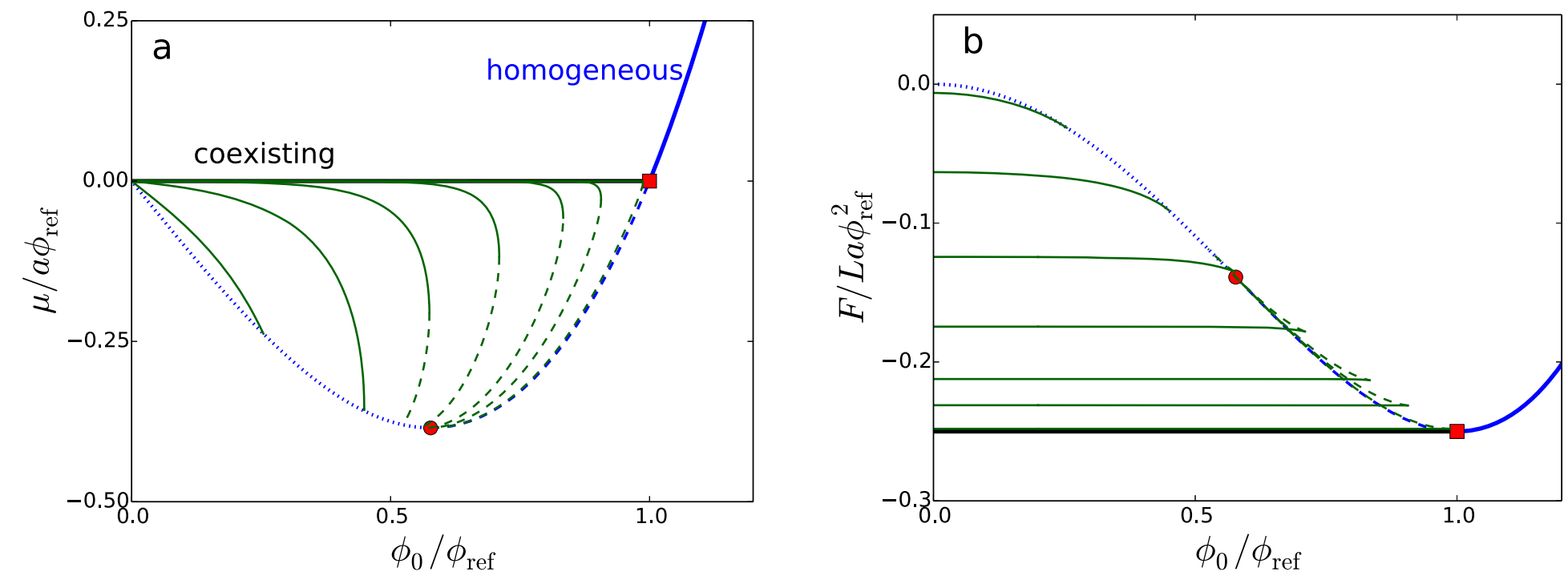

**Figure 4.** (a) The chemical potential $\mu$ and (b) the mean free energy density $F/L$ are shown as functions of the mean concentration $\phi_0 > 0$. The corresponding plots for $\phi_0 < 0$ follow from symmetry (see figure 1). Included are the homogeneous state (blue lines) and the phase-separated state (black line) in the thermodynamic limit, as well as the stable (thin green solid lines) and unstable (thin green dashed lines) steady inhomogeneous states that bifurcate from the homogeneous state at $\phi_L^{n=1}$ for various domain sizes $L/\ell = 7, 10, 15, 25, 50, 10^2, 10^3$. The last is nearly indistinguishable from the thermodynamic limit. Remaining line styles, symbols and definition of $\phi_{\rm ref}$ are as in figure 1.

$$\phi_L^n = \pm\sqrt{\frac{1}{3b}\left[a(T_c - T) - \kappa\left(\frac{2\pi n}{L}\right)^2\right]} \tag{14}$$

that for $L \to \infty$ all converge to $\phi_s$. Numerical continuation allows us to determine the emerging branches of heterogeneous steady states. The most relevant is that with $n = 1$ whose chemical potential and mean free energy are shown for several values of $L$ in figure 4, where the TL is also included.

We now discuss how the character of the bifurcation curves (with $\phi_0$ as the primary control parameter) changes when the secondary control parameter consisting of the domain size $L$ is increased: at the smallest $L$ that allows a heterogeneous state to develop ($L_{\rm min} = 2\pi\ell$ where $\ell = \sqrt{\kappa/a(T_c - T)}$) a pair of supercritical symmetry-breaking pitchfork bifurcations appears in a codimension-2 bifurcation at $\phi_0 = 0$. With increasing $L$ these primary bifurcations move apart, generating in the $\phi_0$ range between them an ever longer branch of heterogeneous states (see e.g. the $L/\ell = 7$ case in figure 4) that represent the lowest energy state at these $\phi_0$ (see figure 4(b)). At $L = L_{\rm ss} = \sqrt{10}\,\pi\ell$ both pitchfork bifurcations become subcritical[8], i.e. the branch of heterogeneous states emerges towards the linearly stable homogeneous states $\phi = \phi_0$. Close to the bifurcation, these heterogeneous states are then linearly unstable and of higher energy than the homogeneous state at identical $\phi_0$, as is clearly visible for curves with $L/\ell > 10$ in figure 4. These unstable states correspond to threshold or nucleation solutions that have to be overcome in order to jump between the linearly stable homogeneous state and the linearly stable heterogeneous states that exist beyond the saddle-node bifurcation at $\phi = \pm\phi_{\rm sn}$ where the branch of heterogeneous states turns around and acquires linear stability. Shortly after turning, the heterogeneous state becomes the global free energy minimum. Typical examples of concentration profiles can be found in the literature; see e.g. figure 3 of [52] and figure 2 of [53].

On further increasing $L$, the primary bifurcation points move further away from $\phi = 0$, ultimately converging on the spinodals $\phi_s$ as $L \to \infty$. At the same time the saddle-node bifurcations move outward and converge on the binodals $\phi_b$. The branch of unstable heterogeneous states becomes longer and ultimately approaches the states represented by a dashed line that correspond to the metastable states in the TL. These states represent a branch of critical nuclei for phase separation. Finally, the branch of stable heterogeneous states between the saddle-nodes at $\pm\phi_{\rm sn}$ becomes increasingly straight and horizontal and converges to the Maxwell line for $L \to \infty$. We emphasize, however, that even in the TL there is no bifurcation between the homogeneous and the heterogeneous state at the binodal and that the branch of unstable nuclei is an intrinsic part of the overall picture.

The overall manner in which the Maxwell construction emerges from the bifurcation scenario in the limit of ever larger domain size is not influenced by the additional branches emerging at $\phi_L^n$ with $n > 1$, since these always correspond to states of higher energy than the $n = 1$ branch and never connect to it. This is because for large $L$ the states on the $n = 1$ branch in a periodic domain consist of a single region where $\phi(\mathbf{x}) \approx \phi_b$ together with a second region where $\phi(\mathbf{x}) \approx -\phi_b$, with two interfaces between them. For $n > 1$ the stationary periodic states consist of a larger number of single phase regions with a correspondingly larger number of interfaces. Note that although the $n > 1$ states never appear as global free energy minima, they may appear as transients in nonequilibrium time simulations. However, this 'simplest possible' picture does not hold in higher dimensions as discussed below.

[8] Weakly nonlinear theory along the lines of sections 4.1 and 5.1 of [50] and [51], respectively, gives $f''f'''' + (f''')^2/3 = 0$ as the condition for the transition from a super- to a subcritical primary bifurcation. This allows one to determine the transition concentration $\phi_{\rm ss} = \phi_{\rm ref}/\sqrt{5}$ and then $L_{\rm ss} = 2\pi/k_{\rm ss}$ via $k_{\rm ss}^2 = -f''(\phi_{\rm ss})$.

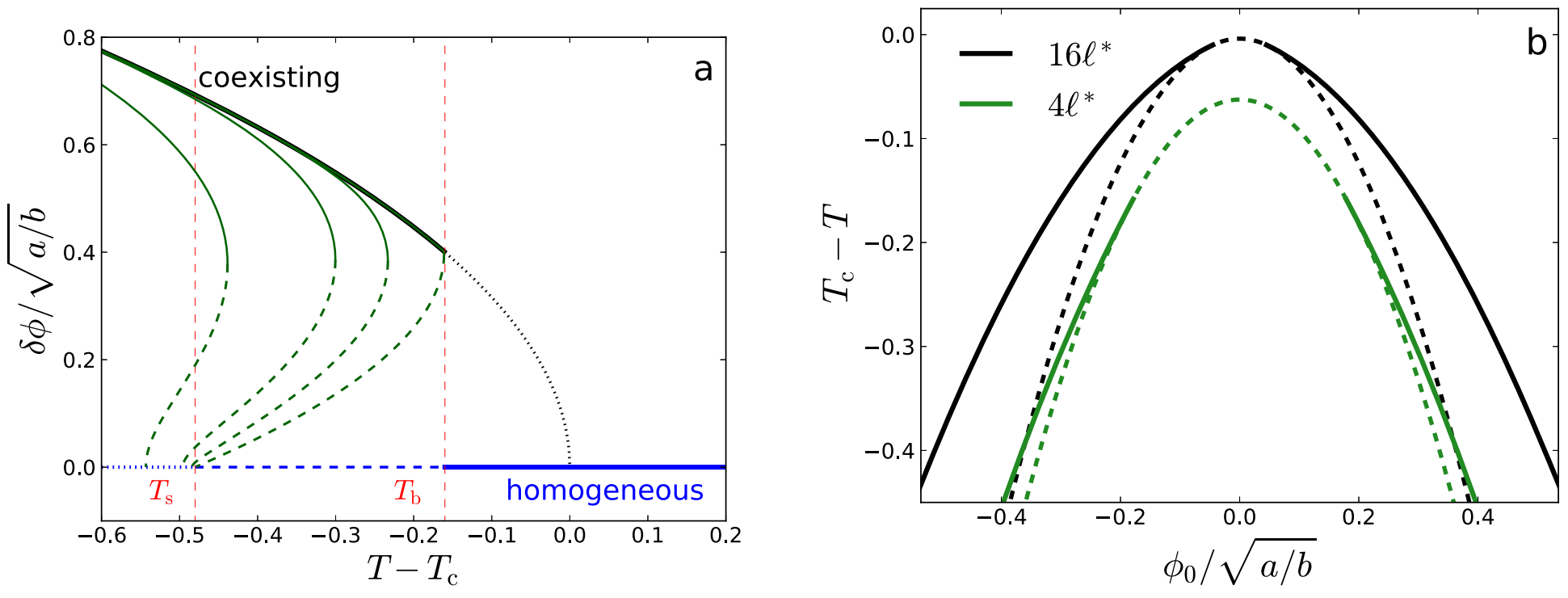

**Figure 5.** Panel (a) shows the order parameter $\delta\phi \equiv (\phi_{max} - \phi_{min})/2$ as a function of the temperature $T$ for $\phi_0 = 0.4$ and domain sizes $L/\ell^\star = 25, 50, 100, 10^4$ (from left to right) where $\ell^\star = \sqrt{\kappa/a}$. The line styles are as in figure 4. The stable (solid) part of the green line for $L/\ell^\star = 10^4$ cannot be distinguished by eye from the demixed state in the thermodynamic limit (solid black line). Panel (b) shows the finite-size equivalent of the phase diagram in figure 3(a) for the two domain sizes $L/\ell^* = 4, 16$ (green and black lines, respectively). Dashed lines indicate the linear stability threshold (finite-size 'spinodal'), while solid lines mark the points where the inhomogeneous state becomes the global energy minimum, if different from the spinodal. This corresponds to a finite-size 'binodal'.

In figure 5(a) we display the value of the order parameter $\delta\phi \equiv (\phi_{max} - \phi_{min})/2$ for various states with $\phi_0 = 0.4$ as the temperature is varied. This is the same as figure 3(b), but now with the heterogeneous states for various values of $L$ also included. We see that as $L \to \infty$ these tend to the curves displayed in figure 3(b). However, we also have a branch of unstable states connecting the two stable branches.

Similarly, one can construct a finite-size equivalent of figure 3(a)—the phase diagram in the TL. First, we need to define what corresponds to the finite-size equivalents of the spinodal and binodal lines. The equivalent of the spinodal corresponds to the linear instability threshold of a homogeneous state of a finite system and is given by equation (14) with $n = 1$. However, for the binodal lines the question is more involved as these correspond to the two coexisting states in the TL which is unaffected by interfaces. Since interfaces are an intrinsic part of (inhomogeneous) states in finite-size systems, the finite-size equivalent of the binodal needs to be defined as a property of these states. We use the concentration values where the inhomogeneous state becomes the global free energy minimum. Figure 5(b) gives the result for two values of the system size. A notable feature is that for finite systems there exists a finite concentration range where the transition is 'second order', together with the corresponding 'tricritical points' where the transition becomes 'first order'. This concentration range increases with decreasing domain size.

### 3.3. Bifurcations in finite domains—demixing in 2D

We now discuss the emergence of the TL for a demixing system in 2D. In contrast to the 1D system discussed in section 3.2, we must now include additional periodic structures in the discussion. As a result, the corresponding bifurcation diagram is richer and the transition to the TL that takes place as the domain size $L \to \infty$ exhibits more interesting features. The inhomogeneous steady states that can arise vary considerably and depend on the domain size, but the most typical ones consist of a circular drop of one phase surrounded by the other or, when $\phi_0$ is closer to zero, of slabs (stripes) connected via the periodic boundary conditions.

Figure 6 shows the chemical potential $\mu$ and rescaled mean free energy density $F/L^2 a\phi_{ref}^2$ for cluster (drop or hole) states as well as for stripe states as a function of the mean concentration $\phi_0$. Square domains of size $L \times L$ are employed and results for two different values of $L$ are given. The stripe states (green lines in figure 6) correspond to the 1D states discussed in section 3.2 that are now extended into the second dimension in a translation-invariant manner. As a result, they have identical bifurcation curves and identical behavior as $L \to \infty$ as the 1D states. However, in 2D new perturbation modes are available and therefore their linear stability properties differ from those in 1D. In particular, the stripe states may now be unstable to transverse perturbations (equivalent to the Plateau–Rayleigh instability of liquid ridges on a solid substrate [54, 55]). Here, this implies that the stripes no longer stabilize at the saddle-node bifurcation at $\phi_{sn}^{stripe}$ where the subcritical part of the branch turns around, as in the 1D case. Instead they become linearly stable at a subsequent secondary pitchfork bifurcation where an unstable branch of transversally modulated stripes (magenta dashed lines in figure 6) emerges subcritically from the branch of translation-invariant stripes.

Additionally, on the square domain considered here, one finds cluster states, consisting of round clusters of phase 2 in phase 1 ('drops', generally for $\phi_0 < 0$) and similar clusters of phase 1 in phase 2 ('holes', generally for $\phi_0 > 0$). These states form a single curve in each bifurcation diagram (red lines in figure 6) and extend between the same primary bifurcation points as the stripe states. As for the stripes, at small $L$ the branch emerges supercritically (not shown) while at larger $L$ the primary bifurcation is subcritical (see, e.g. the case of $L = 4L_c$ in figures 6(a) and (b)). The branch stabilizes at a subsequent saddle-node bifurcation (at $\pm\phi^{\mathrm{cluster}}_{\mathrm{sn}\,1}$) at which it turns around, before losing stability again at a secondary pitchfork bifurcation where the branch of modulated stripes terminates. This shows that the branch of modulated stripes is related to the exchange of stabilities between the stripe and the cluster branch, namely, it corresponds to critical saddle states that have to be overcome in order to transition between these two linearly stable states. On further increasing $L$, the cluster state branch ceases to be monotonic between the two saddle-node bifurcations at $\pm\phi^{\mathrm{cluster}}_{\mathrm{sn}\,1}$ and undergoes a hysteresis bifurcation, at $\phi_0 = 0$, generating two further saddle-node bifurcations (at $\pm\phi^{\mathrm{cluster}}_{\mathrm{sn}\,2}$, see e.g. the case $L = 16L_c$ in figures 6(c) and (d)). This process results in multistability of drop, hole and stripe solutions at small $|\phi_0|$ and in a change in the type of the lowest energy state with changing $\phi_0$ as indicated by the thick solid lines in figure 6. Images of such states can be found, e.g. in figure 2 of [56], figure 2 of [57] and figures 8 and 10 of [58]. Note that in the $\mu(\phi_0)$ diagram the branch segments between $\pm\phi^{\mathrm{cluster}}_{\mathrm{sn}\,1}$ and $\pm\phi^{\mathrm{cluster}}_{\mathrm{sn}\,2}$ become almost horizontal near $\pm\phi^{\mathrm{cluster}}_{\mathrm{sn}\,2}$. These plateaus approach each other as $L$ increases and we use the quantity $(\mu(\phi^{\mathrm{cluster}}_{\mathrm{sn}\,2}) - \mu(-\phi^{\mathrm{cluster}}_{\mathrm{sn}\,2}))/2 \equiv \mu^{\mathrm{cluster}}_{\mathrm{sn}\,2}$ to illustrate this tendency—see figure 7.

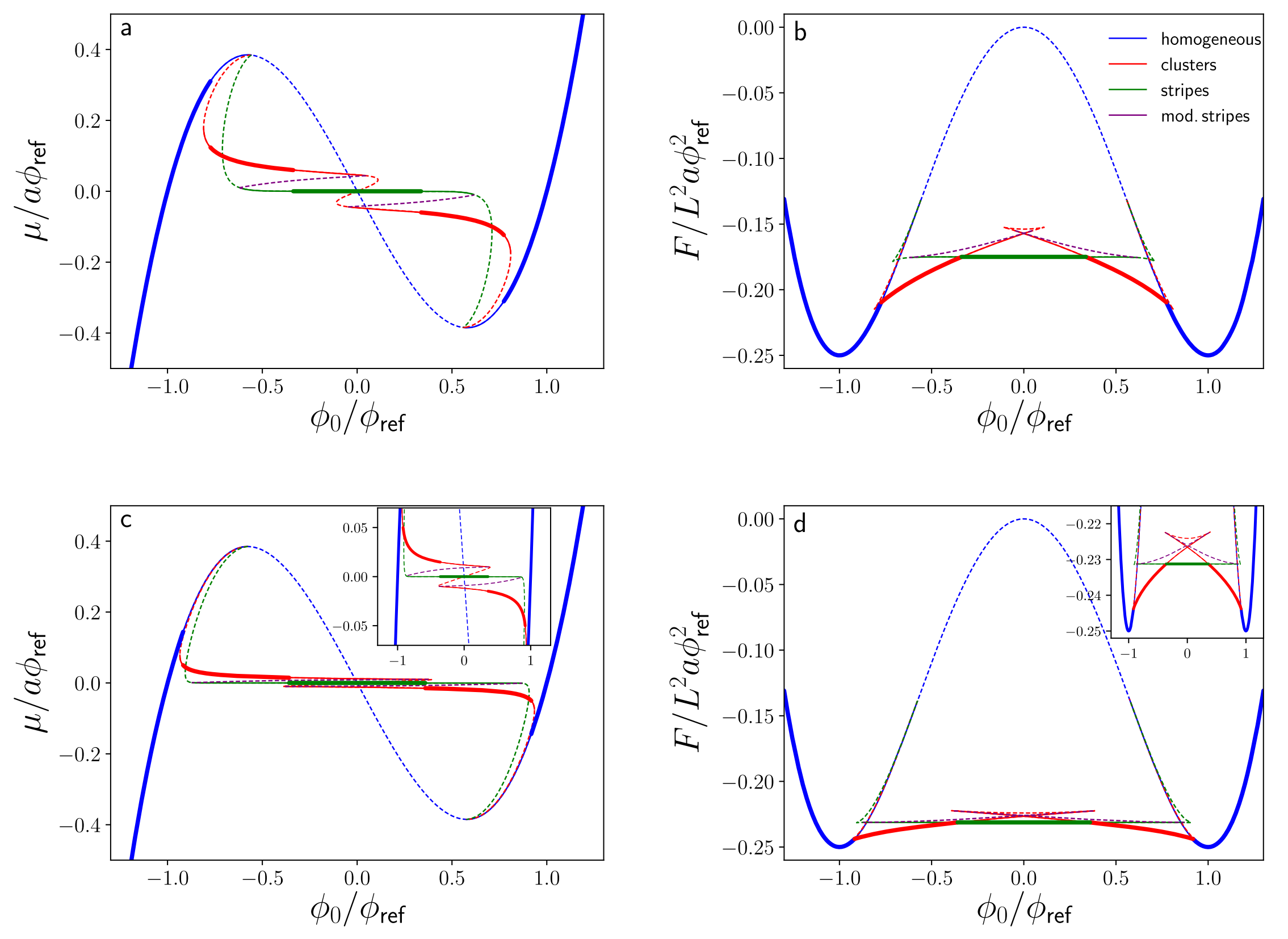

**Figure 6.** Bifurcation diagrams of steady states of the Cahn–Hilliard equation in 2D. The various colors indicate different states as given in the legend. The first and second row show (a), (c) the chemical potential $\mu$ and (b), (d) the rescaled mean free energy density $F/L^2a\phi^2_{\mathrm{ref}}$ as a function of the mean concentration $\phi_0$ for square domains of size $L \times L$ with $L = 4L_c$ and $L = 16L_c$, respectively, where $L_c = 2\pi$. Linearly stable (unstable) states are indicated by thin solid (dashed) lines. The thick solid lines denote the states of minimum free energy at each value of $\phi_0$.

With increasing domain size $L$ the saddle-node bifurcations at $\pm\phi^{\mathrm{cluster}}_{\mathrm{sn}\,1}$ on the cluster branch and the corresponding bifurcations at $\pm\phi^{\mathrm{stripe}}_{\mathrm{sn}}$ on the stripe branch gradually move outwards and approach the branch of homogeneous states at the Maxwell point. The saddle-node bifurcations at $\pm\phi^{\mathrm{cluster}}_{\mathrm{sn}\,2}$ on the cluster branch likewise move outwards towards larger $|\phi_0|$. As $L \to \infty$ they approach $\pm\phi^{\infty}_{\mathrm{sn}\,2}/\phi_{\mathrm{b}} = \pm(\pi/2 - 1) \approx \pm0.571$, a prediction that follows from the sharp interface limit[9]. During this process the two plateaus in $\mu$ become longer and their difference in chemical potential, $\Delta\mu \equiv 2\mu^{\mathrm{cluster}}_{\mathrm{sn}\,2}$, decreases as $L^{-1}$. The convergence of $\pm\phi^{\mathrm{stripe}}_{\mathrm{sn}}$, $\pm\phi^{\mathrm{cluster}}_{\mathrm{sn}\,1}$ and $\pm\phi^{\mathrm{cluster}}_{\mathrm{sn}\,2}$ as well as the way all $\mu_{\mathrm{sn}}$ approach zero with increasing $L$ is illustrated in the log–log plots of figure 7. There we see that $\mu^{\mathrm{cluster}}_{\mathrm{sn}\,1}$ decreases as $L^{-2/3}$.

[9] In the sharp interface limit we assume that regions of $\phi = \phi_{\mathrm{b}}$ and $\phi = -\phi_{\mathrm{b}}$ are separated by a sharp interface. On a domain $L \times L$, a circular cluster of radius $R$ of $\phi_{\mathrm{b}}$ in a background of $-\phi_{\mathrm{b}}$ can exist for $0 \leqslant R \leqslant L/2$, i.e. up to $\phi_0/\phi_{\mathrm{b}} = [\pi R^2 - (L^2 - \pi R^2)]/L^2 = \pi/2 - 1$.

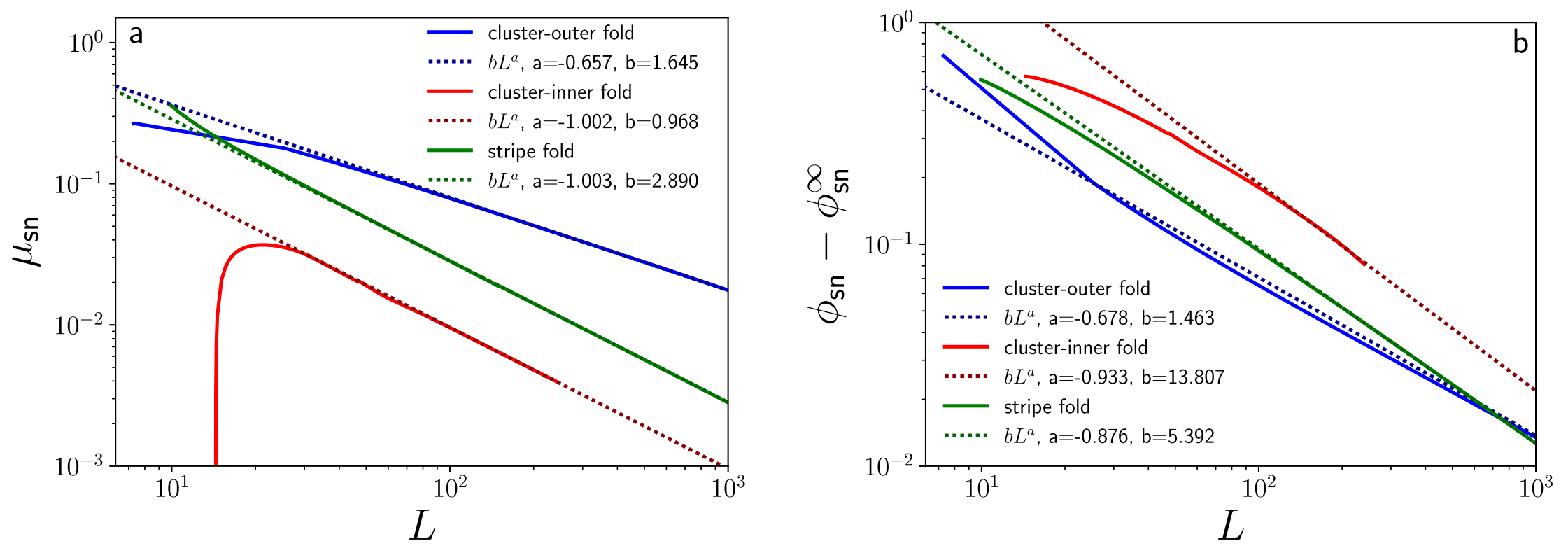

**Figure 7.** Domain-size dependence of various characteristic quantities showing how the Maxwell construction comes into being as $L$ increases. The log–log plots indicate that (a) the chemical potentials at the saddle-node bifurcations $\mu_{\mathrm{sn}\,1}^{\mathrm{cluster}}$, $\mu_{\mathrm{sn}\,2}^{\mathrm{cluster}}$ and $\mu_{\mathrm{sn}}^{\mathrm{stripe}}$, and (b) the corresponding differences in concentrations $\phi_{\mathrm{sn}\,2}^{\mathrm{cluster}} - \phi_{\mathrm{sn}\,2}^{\infty}$, $\phi_{\mathrm{sn}\,1}^{\mathrm{cluster}} - \phi_{\mathrm{b}}$ and $\phi_{\mathrm{sn}}^{\mathrm{stripe}} - \phi_{\mathrm{b}}$ approach zero in the form of power laws as $L \to \infty$ (see legend) and describe how the three overlapping $\mu$ plateaus corresponding to drop, hole and stripe states approach one another to form a single Maxwell line in the limit.

Note that the stripe state (parallel slabs of the two coexisting phases) is the state of lowest free energy state in a region centered on $\phi_0 = 0$, while for larger $|\phi_0|$ the cluster state has the lowest free energy (see figures 6(b) and (d)). The sharp interface limit indicates that in the limit $L \to \infty$ the stripe state [cluster state] represents the minimum energy state provided $|\phi_0/\phi_\mathrm{b}| < 2/\pi - 1 \approx 0.363$ [$1 > |\phi_0/\phi_\mathrm{b}| > 2/\pi - 1$][10]. The values of $\phi_{\mathrm{sn}} - \phi_{\mathrm{sn}}^{\infty}$ also decrease as power laws as $L \to \infty$, apparently in the same manner as $\mu_{\mathrm{sn}}$, as revealed in figure 7(b).

In this section we have seen that close to the primary bifurcation the emerging curves of 2D inhomogeneous states behave much like in the 1D case. However, in the region around $\phi_0 = 0$ the behavior differs strongly as more states are accessible to the system in 2D, resulting in a different convergence pathway to the Maxwell line: several $\mu$ plateaus develop that converge to the single horizontal line of the Maxwell construction only in the TL. However, different regions on this line continue to represent different states.

## 4. Crystallization

### 4.1. Phase behavior

As in section 3.1 for liquid–liquid demixing, we first review the crystallization phase behavior described by the free energy (6). Minimizing $F[\phi]$ at fixed mean density $\phi_0$ gives

$$r\phi + (q^2 + \Delta)^2\phi + \phi^3 = \mu, \qquad (15)$$

where $\mu$ represents the chemical potential that determines the mean value of the order parameter $\phi_0$, here defined as $\phi_0 = V^{-1}\int_V \phi \, \mathrm{d}\mathbf{x}$, where $V$ is the volume of the domain.

The liquid state is homogeneous and the corresponding spatially uniform solution of equation (15) exists for all $\phi_0$: given $\mu$, $\phi_0$ solves $(r + q^4)\phi_0 + \phi_0^3 = \mu$. When $r + 3\phi_0^2 = 0$ this state becomes linearly unstable to perturbations with wave number $q$, i.e. the spinodal is given by $\phi_\mathrm{s} = \sqrt{-r/3}$, and the critical point is at $(r_\mathrm{c}, \phi_\mathrm{c}) = (0, 0)$ (maximum of the red dashed line in figure 8). In the $r$ range directly below the critical point, the crystallization phase transition is predicted by the PFC model to be of second order. The transition is of first order only below the 1D tricritical point situated at $(r_{\mathrm{tri}},\ \phi_{\mathrm{tri}}) = (-9/38,\ \pm\sqrt{3/38}\,)$ [20]. Since freezing is in reality a first order transition [59], the PFC model is not correct for $r > r_{\mathrm{tri}}$. This is a consequence of the approximations made in deriving the model [38].

At $(r_{\mathrm{tri}}, \phi_{\mathrm{tri}})$ the binodals emerge (black solid lines in figure 8) that limit the coexistence region and specify the densities $\phi_\mathrm{b}^{\mathrm{li}}$ and $\phi_\mathrm{b}^{\mathrm{cr}}$ of the liquid and crystalline states that coexist at a given undercooling $r$, i.e. the states with the same chemical potential and pressure (i.e. the same grand potential density). The binodals can either be calculated for particular spatial periods or wavelengths of the crystalline structure or for an infinite domain. In the latter case, the wavelength of the crystalline state is that which minimizes the free energy. The binodals in figure 8 are calculated at fixed structural length corresponding to the critical wave number $L_\mathrm{c} = 2\pi/q$. Note that in the following we always use $q = 1$, i.e. $L_\mathrm{c} = 2\pi$. The results of the two approaches cannot in general be distinguished on scales such as that used in figure 8. As for the Cahn–Hilliard equation, the binodal lines are determined using numerical path continuation as done for the 1D case in [20]: first, we identify the coexisting values of the densities $\phi_b^{li}$ and $\phi_b^{cr}$ at some particular undercooling $r$. We then continue these values as a function of $r$.

[10] In the sharp interface limit, the interface energy of a stripe state in a $L \times L$ domain is $\sim 2L$, independently of $\phi_0$. For the circular cluster state the interface energy is $\sim 2\pi R$ where the radius $R$ depends on $\phi_0$. Averaging the density for a fully decomposed state with $\phi = \pm\phi_\mathrm{b}$ one obtains $\phi_0/\phi_\mathrm{b} = \pi r^2 - (1 - \pi r^2)$ where $r = R/L$. This gives $R = L\sqrt{(\phi_0/\phi_\mathrm{b} + 1)/2\pi}$, implying that the interface energy of the cluster state is lower than that of the stripe state when $\sqrt{\pi(\phi_0/\phi_\mathrm{b} + 1)/2} < 1$, i.e. when $|\phi_0/\phi_\mathrm{b}| > 2/\pi - 1$.

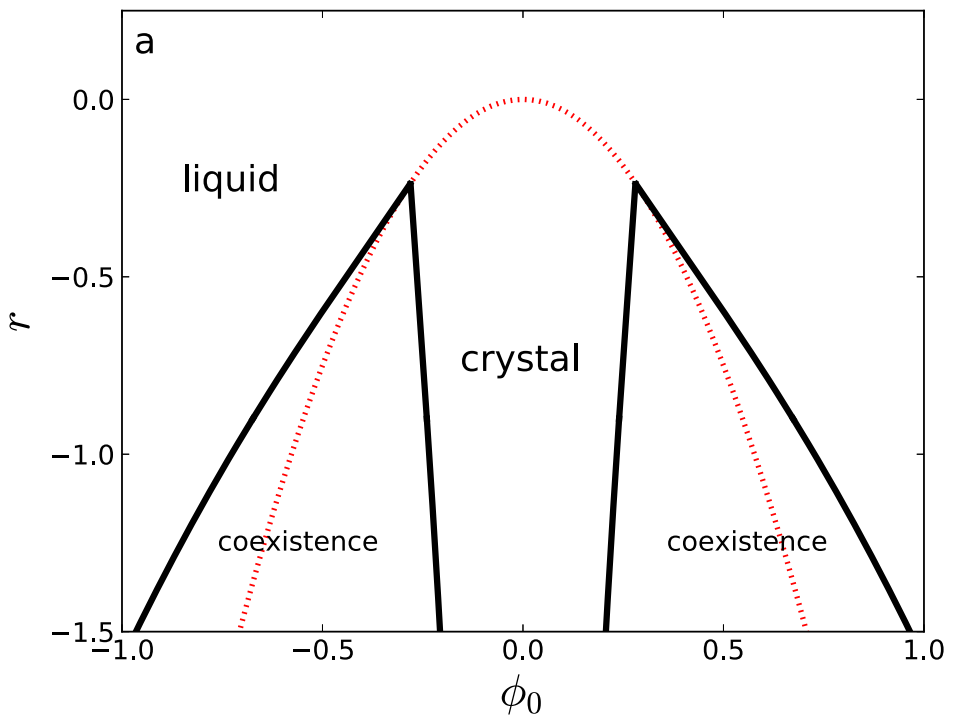

**Figure 8.** Phase diagram of the PFC model in the plane spanned by the mean concentration $\phi_0$ and the undercooling $r$ on (a) one-dimensional (1D) and (b) two-dimensional (2D) domains. Shown are the spinodal line (dotted line), where the liquid state becomes linearly unstable, and the binodal lines (solid lines) indicating the liquid (outer solid lines) and crystalline (inner solid lines) states that coexist at fixed $r$. In the 2D case, only the binodals for the transition from liquid (solid line on the left) to hexagonal (solid line on the right) crystal are shown.

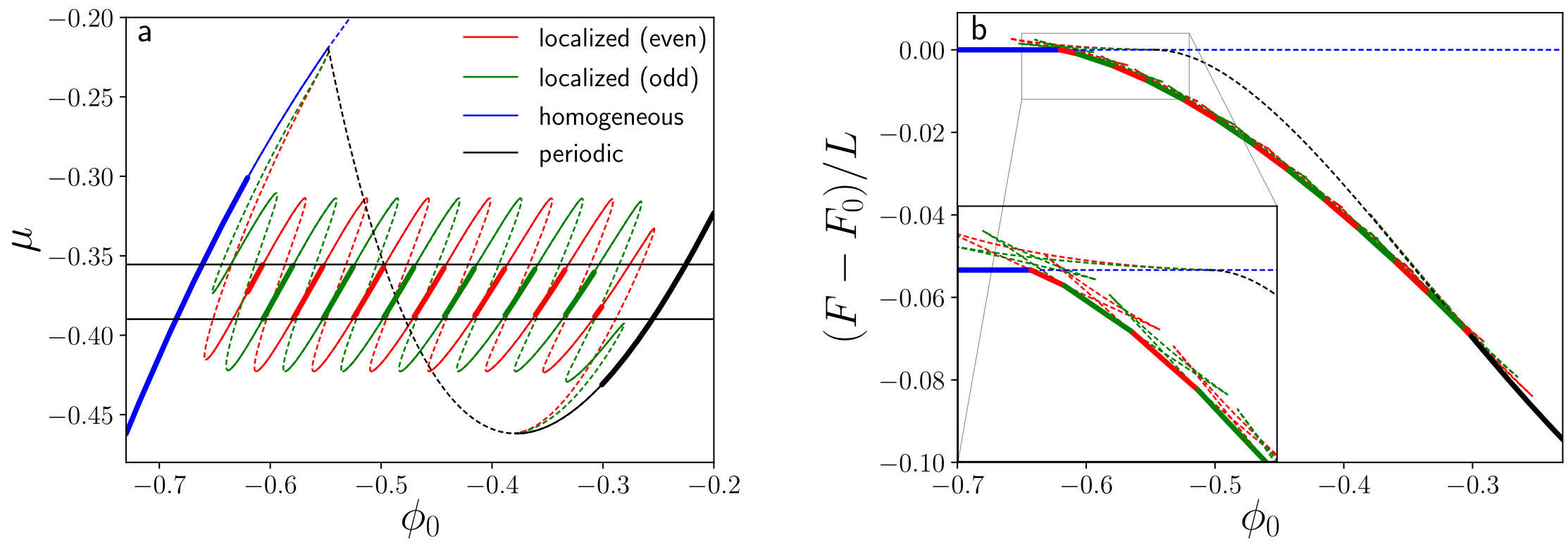

**Figure 9.** Bifurcation diagram for the 1D PFC system in a domain of size $L = 16L_c$ showing crystallites (spatially localized states) with even (red line) and odd (green line) numbers of bumps, the liquid (homogeneous) state (blue line) and the crystalline (periodic) state (black line). Linearly stable and unstable states are indicated by solid and dashed lines, respectively. All are characterized by (a) their chemical potential $\mu$ and (b) their relative mean free energy density $(F - F_0)/L$ where $F_0 = F[\phi_0]$. Global minima (determined from (b), for details see inset) are indicated by thick line segments. When connected up, the global minima corresponding to localized states that lie on a curve with piecewise constant slope: a zigzag curve. The width $\Delta\mu$ of this zigzag region is indicated in (a) by a pair of horizontal lines.

Figure 9 shows the chemical potential $\mu$ and the mean free energy density $F/L$ of the 1D liquid and the space-filling crystalline states at fixed $r$ as a function of $\phi_0$ using, respectively, blue and black lines[11]. Equation (15) also possesses solutions corresponding to finite size portions of crystal that coexist with a liquid background. These crystallite 'localized states' are present in the coexistence region (and slightly outside) and are discussed in greater detail in [20] (see their figures 3 and 5 for typical solution profiles). These states exhibit the phenomenon of slanted snaking [13, 61, 62] but take the form of standard homoclinic snaking when the order parameter $\phi_0$ is plotted as a function of the chemical potential, a property that is expected to carry over to other models with a conserved order parameter. See the conclusion of [20, 26] for further discussion. Note that figure 9 highlights the states representing global energy minima using thick line segments, an approach also taken for other systems showing multiple steady states, as done, e.g. in [14] for various buckling states.

In 2D the phase diagram of the PFC system is much richer (see figure 10 of [20]). In addition to the uniform or liquid state, the PFC model now exhibits stripe-like states, and two distinct periodic states with hexagonal coordination, one of which consists of density maxima on a hexagonal lattice (referred to as bumps, i.e. the crystal state) and the other with similarly arranged density minima (referred to as holes). As a result, the system may exhibit phase coexistence between the uniform state and the bump state, between the bump state and stripes, between stripes and holes and between holes and the uniform state. Here, we only consider the first scenario arising from thermodynamic coexistence between the liquid (homogeneous) state and the crystalline (bump) state, and the spatially localized structures associated with this coexistence. Figure 8(b) shows the corresponding part of the phase diagram, with the spinodal line indicating the onset of linear stability of the liquid state and the binodals specifying the coexisting liquid and hexagonal crystal states at a given $r$. As in 1D, the coexistence region is associated with the presence of 2D localized crystallites.

[11] Similar diagrams for varying $r$ at fixed $\phi_0$ for a related PFC model can be found in [60].

In the next two sections we show that the localized states present in both 1D and 2D are intimately related to the emergence of the Maxwell construction in the TL for the liquid to crystal phase transition. The results represent a third way in which the Maxwell construction is approached as $L \to \infty$. In both cases, 1D and 2D, we use the values of the chemical potential at coexistence as reference values.

### 4.2. Bifurcations in finite domains: crystallization in 1D

The properties of the liquid (homogeneous) and crystalline (periodic) bump states and of the localized structures associated with their coexistence are summarized in figure 9 in terms of a bifurcation diagram for a relatively small domain of $L = 16L_c$. The figure shows (a) the chemical potential $\mu$ and (b) the relative mean free energy density $(F–F_0)/L$ where $F_0 = F[\phi_0]$ as a function of $\phi_0$, for $r = -0.9$. Thin solid and dashed lines are employed for linearly stable and unstable states, respectively. The figure also indicates, using thick solid lines, the thermodynamically stable state for each value of the mean concentration $\phi_0$, i.e. the global free energy minimum. These states are identified in figure 9(b). For alternative representations using the norm or grand potential as order parameters and typical solution profiles, see [20].

Figure 9 shows that the periodic (or domain-filling crystalline) state (black line) bifurcates in the direction of decreasing stability of the uniform state $\phi = \phi_0$ (blue line), i.e. supercritically. This crystalline state loses stability at small amplitude to a pair of spatially localized structures (red and green lines) resembling crystalline states of finite length embedded in the background uniform liquid state. These states are distinguished by their behavior at $x = 0$, i.e. at the center of the pattern, with the states in red having a maximum at $x = 0$ and those in blue having a minimum at $x = 0$, and both exhibit snaking. This behavior in turn implies the presence of interconnecting branches (resembling rungs on a ladder) of unstable asymmetric structures that are computed in [20] but not shown here. The figure indicates the linear stability properties of each state shown, with thin solid lines indicating linearly stable states (or local free energy minima) and thin dashed lines indicating unstable states. Note that the localized states coexist with the stable uniform state, a possibility that only arises because of mass conservation [63], and that they represent the global free energy minimum over a large part of their range of existence (figure 9(b)). The inset of figure 9(b) indicates that the transition between successive global minima occurs via swallow-tail-like structures.

Figure 9(a) shows that the localized states corresponding to global free energy minima are clustered within a band of width $\Delta\mu \approx 0.031\,08$ within the snaking region, until superseded by the periodic state at large $\phi_0$. Figure 10(a) displays only these global free energy minima for various domain sizes $L$. Connecting adjacent branches of such minima generates a zigzag curve, and figure 10(a) shows such zigzag curves for eight different domain sizes ranging from $L = 6L_c$ to $512L_c$. We see that the slopes of the slanted segments of the curves are all the same and independent of the domain size $L$. However, the length of these segments decreases with increasing $L$ leading to a corresponding increase in the number $N(L)$ of slanted segments. Figure 10(b) shows that the resulting stability interval $\Delta\mu$ follows, with excellent accuracy, the power law $\Delta\mu \sim aL^b$ with exponent $b \sim -1$. This exponent is consistent with the observation that the slope of the slanted portions of the curve is independent of $L$ and moreover that the number of oscillations in each snaking curve is proportional to $L$. The latter is in turn a consequence of the fact that each oscillation in the curve results in the addition of the same wavelength on either side of the localized state as the structure grows.

We conclude that in the limit $L \to \infty$ the stability interval shrinks to zero and conjecture that in the limit the resulting curve is nowhere differentiable. We also confirm that the location of the limiting chemical potential corresponds precisely to the Maxwell construction for these parameter values. Thus, from this point of view the Maxwell construction involves a continuum of localized structures, all coexisting at the same value of the chemical potential $\mu$.

### 4.3. Bifurcations in finite domains: crystallization in 2D

We next consider the case of two-dimensional domains focusing on the transition between a liquid state and a hexagonal crystalline state. For this purpose we use hexagonal domains with specific side lengths as physical domains $\mathcal{D}$. As explained in section 2.3, numerical continuation is then applied on a domain $\mathcal{D}^{\mathrm{num}}$ that corresponds to one twelfth of $\mathcal{D}$ with appropriate boundary conditions reflecting the symmetries of the states considered. Employing other domain shapes and boundary conditions corresponding to different symmetries can affect aspects of the results. For instance, a 'wrong' domain shape can rule out the existence of the periodic hexagonal state and bifurcations may become imperfect. However, if the domain is sufficiently large one always finds the snaking branches of hexagonal patches discussed below as long as these are small compared to the domain. The only part of the bifurcation structure that is affected by the domain corresponds to the transition from the crystalline patch state to a periodic crystalline state, and reflects the interaction of the patches with the boundary. For hexagonal domains of different side lengths $L_h$, figure 11 displays the corresponding bifurcation diagrams of steady homogeneous, periodic and localized states for $r = -0.9$. Examples of localized states with hexagonal coordination are presented in figure 12. At first sight, the diagrams in figure 11, which are in terms of the chemical potential and the free energy density, have a similar appearance to those for the 1D case displayed in figure 9. However, they differ in several significant aspects: following the stable homogeneous state in the direction of increasing density, we see that this time it becomes unstable in a subcritical pitchfork bifurcation where a solution branch of defect-free domain-filling hexagonal patterns emerges (black dashed curve). A branch of nearly rotationally invariant localized target-like solutions (dark green dashed curve) emerges from this branch at very small amplitude, even though the domain does not have this symmetry (see figure 12(a)). The target-like structure grows in radius along the branch, although only part of it is shown in figure 11. Although nominally axisymmetric the target structures do in fact reflect the symmetry of the domain which is responsible for the presence of a very slight $D_6$ deformation of this state.

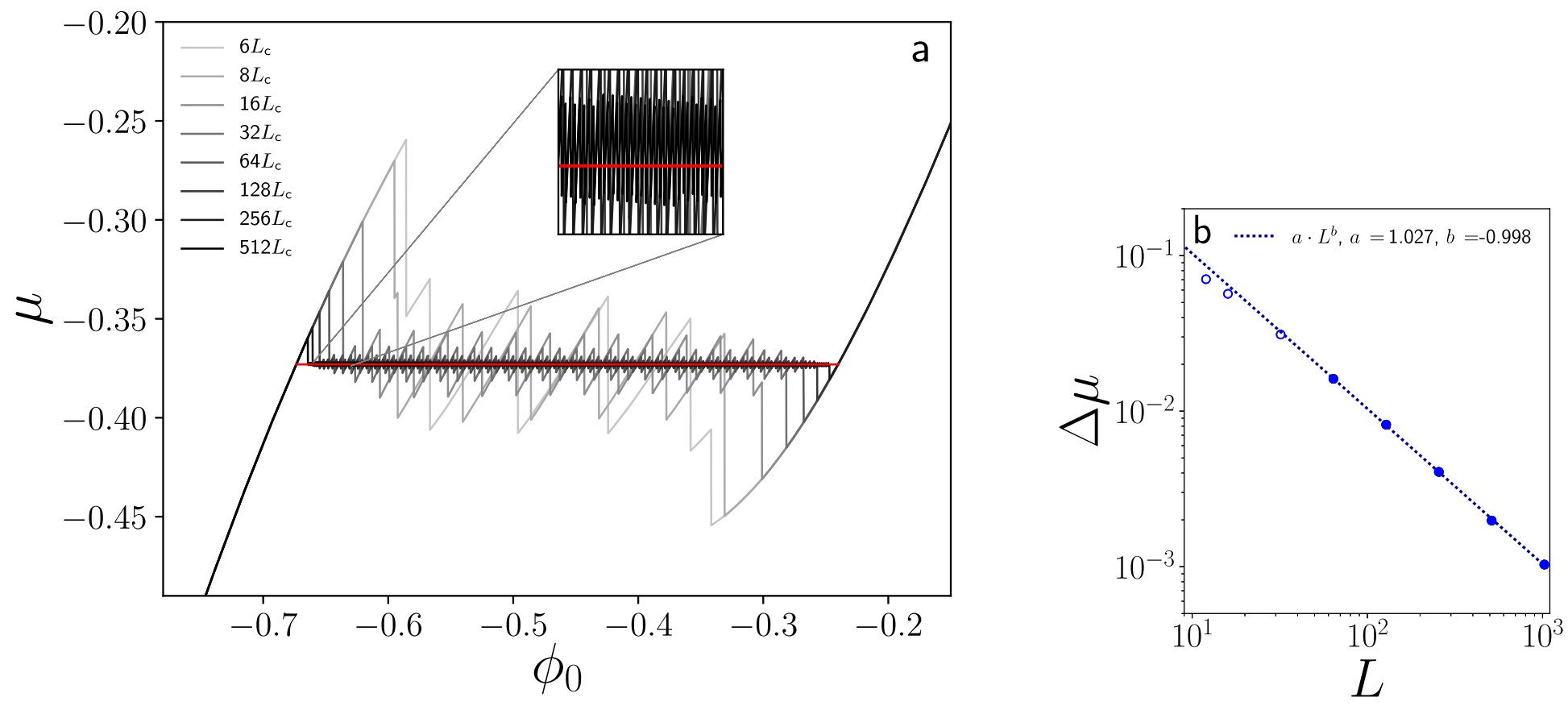

**Figure 10.** (a) As in figure 9(a) but only showing the zigzag curves, i.e. the global minima of the free energy, for domain sizes from $L = 6L_c$ to $512L_c$ with $L_c = 2\pi$, as given in the legend. The horizontal red line limited by the binodal values $\phi_b^{li}$ to the left and $\phi_b^{cr}$ to the right indicates the chemical potential at coexistence, i.e. the Maxwell construction, as obtained in section 4.1 (see figure 8(a)). As $L$ increases, the oscillation period and the width $\Delta\mu$ of the zigzag curve both decrease. The leftmost peak where the homogeneous state first loses stability to localized states likewise decreases significantly. The corresponding peak on the right where the localized states reach the periodic state is very sensitive to numerical precision. For $L = 512L_c$ only part of the curve is calculated—its structure is visible in the inset. Panel (b) gives the corresponding widths $\Delta\mu$ of the zigzag region as a function of the domain length $L$ showing that $\Delta\mu \sim L^b$ with $b \approx -1$, as expected from the fact that the slope of the slanted portions of the curve is independent of $L$ and the number of oscillations is proportional to $L$. The fit neglects the smallest values for $L$ where the slopes vary too much for the above argument to apply.

The branch of localized hexagonal patches (red curve) bifurcates in a tertiary steady-state bifurcation from the target pattern branch. The bifurcation is a $D_6$-equivariant pitchfork bifurcation and so breaks the (nominal) axisymmetry of the target state. However, the above-mentioned slight $D_6$ deformation of the target structures makes the bifurcation slightly imperfect as can be ascertained from continuation runs with a smaller stepsize.

The states presented in figure 12 (with the exception of the first profile) all lie on this latter emerging branch and show that as one follows the branch the hexagonal patch gradually increases in size. This process occurs via the addition of new bumps at preferred locations along each side and is followed by the addition of further bumps on either side until each row is complete. The first two profiles in the second row show some of the intermediate states on the branch segment extending from a 4 bumps-on-a-side patch to a 5 bumps-on-a-side patch. Hexagonal patches in the nonconserved Swift–Hohenberg equation grow in the same manner [64]. The appearance of each new cell is associated with a fold in the snaking branch, i.e. as the patch increases in size the density of the branches in the bifurcation diagram increases.

Inspection of figures 11(a) and (c) shows that the localized hexagonal patches corresponding to global minima of the free energy (figures 11(b) and (d)) are again confined to a zigzag curve straddling the Maxwell point for the unbounded system. These zigzag curves are shown in figure 13 for $L_h = 4L_c^h$, $6L_c^h$, $8L_c^h$ and $16L_c^h$ and show behavior that is, in principle, similar to that in figure 10(b). However, the resulting structure is more rugged and irregular, and the width $\Delta\mu$ has to be carefully defined as the individual inclined straight segments are not all centered about the coexistence value. We therefore determine the maximum and minimum $\mu$ value for each inclined segment and average them to obtain $\bar{\mu}_{max}$ and $\bar{\mu}_{min}$, thereby obtaining $\Delta\mu \equiv \bar{\mu}_{max} - \bar{\mu}_{min}$. Figure 14(b) shows the resulting widths of the region of thermodynamically stable patches $\Delta\mu$ as a function of the domain size $L_h$. Based on the following estimate, we expect the dependence to follow the power law $\Delta\mu \sim L^{-2}$. To complete a hexagon side of length $NL_c^h$ the bifurcation curve passes through $\frac{N+1}{2}$ (rounded to an integer) folds. To create the whole hexagonal patch starting from a single central peak takes $1 + 2 + 2 + 3 + 3 + 4 + 4 + 5 + 5 + \ldots + \frac{N+1}{2} \approx \frac{N+1}{2}(N+3)$ folds. In terms of the side length, the number of folds is proportional to $L_h^2 + 4L_h$. Next, we note that the number of zigzag segments that correspond to a global minimum for a given $\phi_0$ is proportional to the number of folds and assume that their slope is constant. Then, $\Delta\mu$ is inversely proportional to the number of folds, i.e. $\Delta\mu \sim (L_h^2 + 4L_h)^{-1}$. Therefore, for sufficiently large domains one expects to find the power law $\Delta\mu \sim L_h^{-2}$. However, in our calculations for the hexagonal patches we are only able to go to a maximum side length of $32L_c^h$. Despite this limitation the observed dependence of $\Delta\mu$ on $L$ is well described by the full expression derive above, as illustrated in figure 14(b).

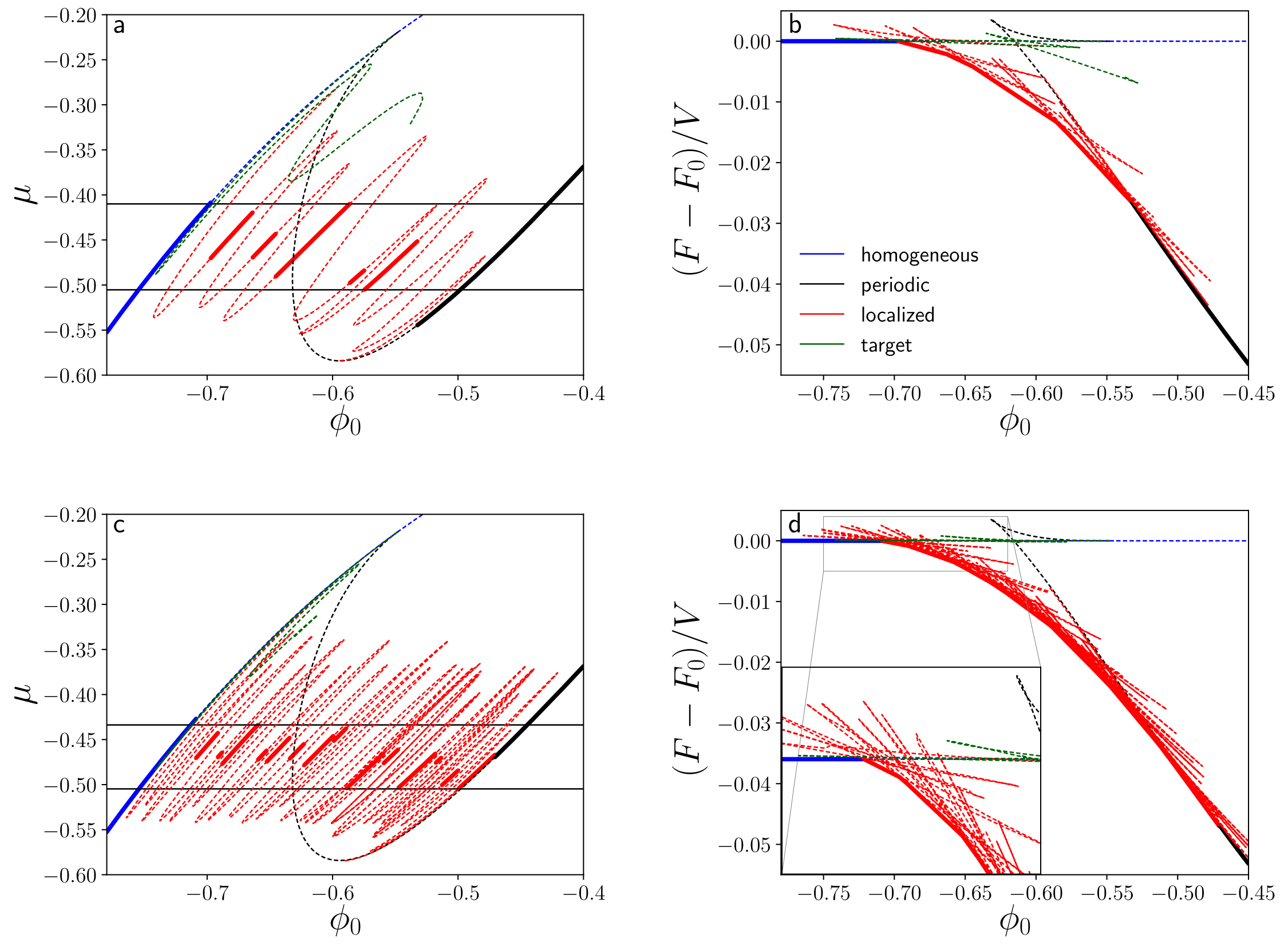

**Figure 11.** Bifurcation diagrams of steady states of the PFC equation in 2D for hexagonal domains of side length $L_h$ with (a), (b) $L_h = 4L_c^h$ and (c), (d) $L_h = 8L_c^h$, where $L_c^h = 4\pi/\sqrt{3}$. Different colors indicate different states, as given in the legend. (a), (c) The chemical potential $\mu$ and (b), (d) the relative mean free energy density $(F - F_0)/A$ where $A = 3/2\sqrt{3}\,L_h^2$, both as a function of the mean concentration $\phi_0$. Global minima (determined from panels (b) and (d)) are indicated by thick line segments. In each case the width $\Delta\mu$ of the resulting zigzag region is indicated by a pair of horizontal lines. Typical profiles for $L_h = 8L_c^h$ are shown in figure 12.

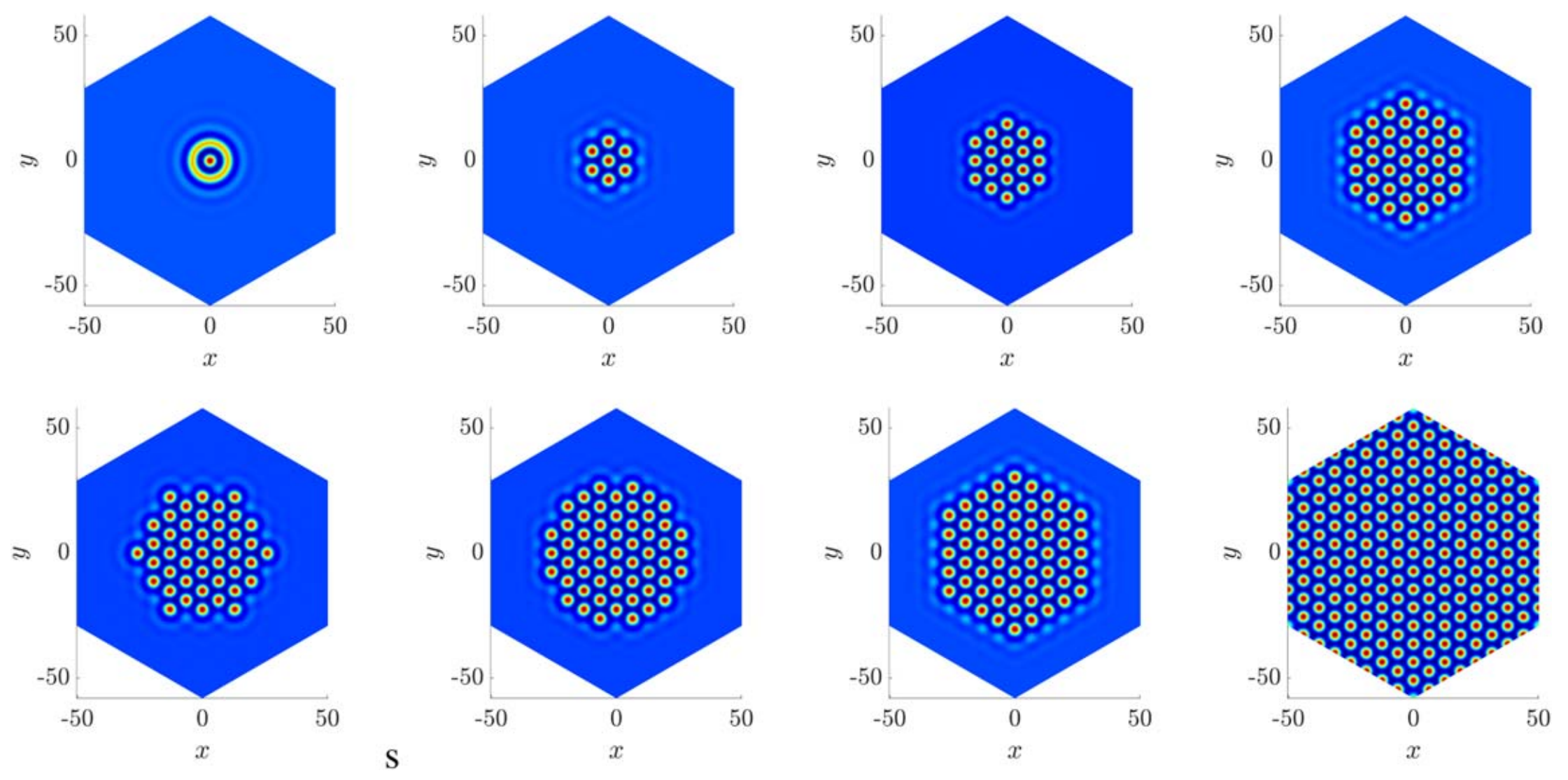

**Figure 12.** Typical density profiles of localized states on a hexagonal domain of side length $L_h = 8L_c^h$ (see the bifurcation diagrams in figures 11(c) and (d)). In the first row we show a single target state and three localized hexagonal patch solutions with completely filled rows. The parameters from left to right are $\mu = -0.3566$, $\phi_0 = -0.6475$; $\mu = -0.3319$, $\phi_0 = -0.6330$; $\mu = -0.4601$, $\phi_0 = -0.7024$ and $\mu = -0.3397$, $\phi_0 = -0.5920$. The second row shows localized states where an additional row is developing through intermediate states. The final picture gives a nearly domain-filling hexagonal pattern. From left to right, the parameters are $\mu = -0.4010$, $\phi_0 = -0.6277$; $\mu = -0.4053$, $\phi_0 = -0.6133$; $\mu = -0.3414$, $\phi_0 = -0.5561$ and $\mu = -0.4952$, $\phi_0 = -0.4916$.

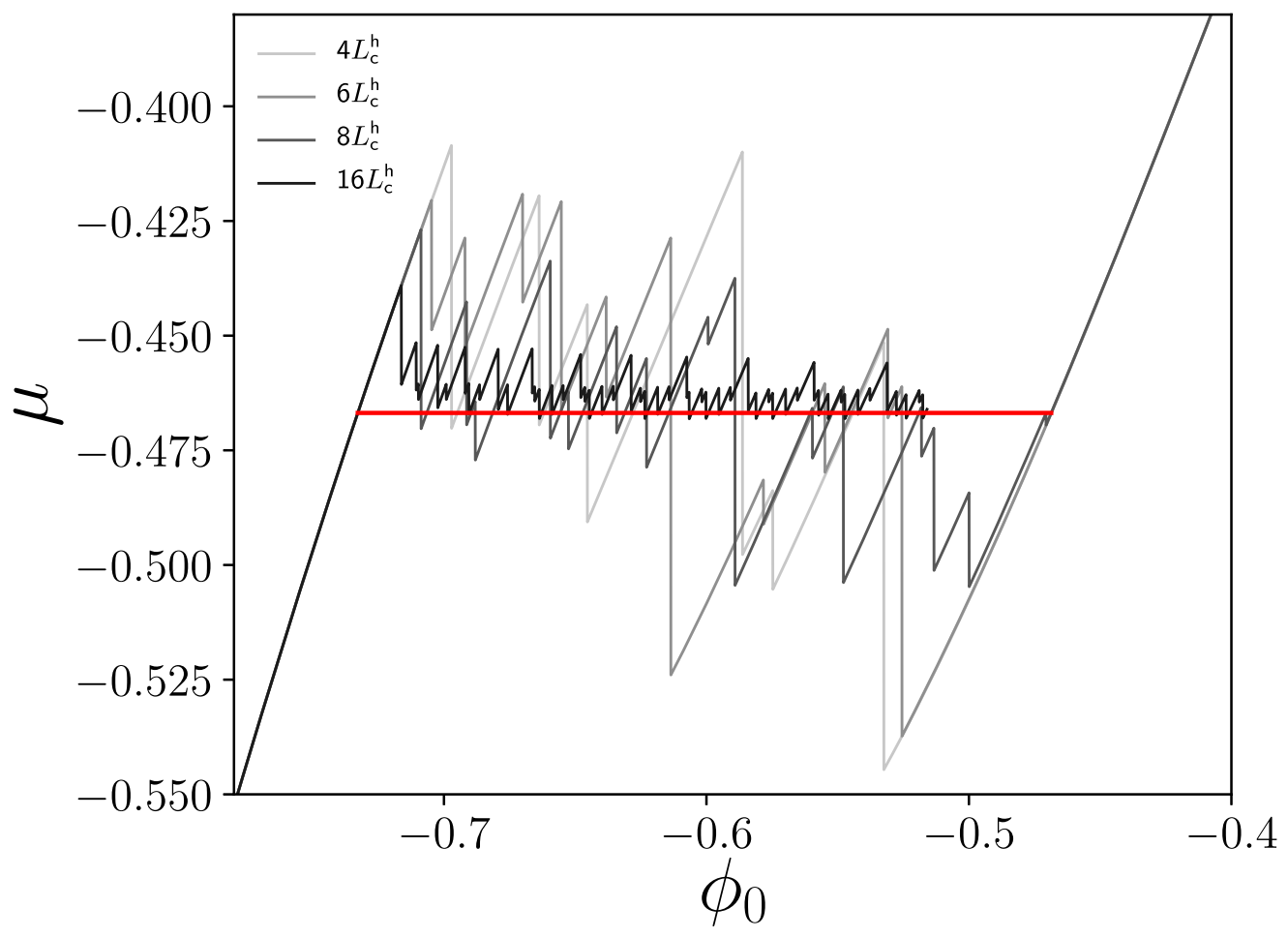

**Figure 13.** As in figures 11(a) and (c) but showing the zigzag curves consisting only of the global minima of the free energy density, for hexagonal patches in domains with $L_h = 4L_c^h$ to $16L_c^h$ as given in the legend. As $L_h$ increases the length of the sloped segments and the width $\Delta\mu$ of the zigzag curves both decrease. The curve for $16L_c^h$ ends once boundary effects complicate the calculation. The horizontal red line indicates the Maxwell construction, as obtained in section 4.1 (see figure 8(b)).

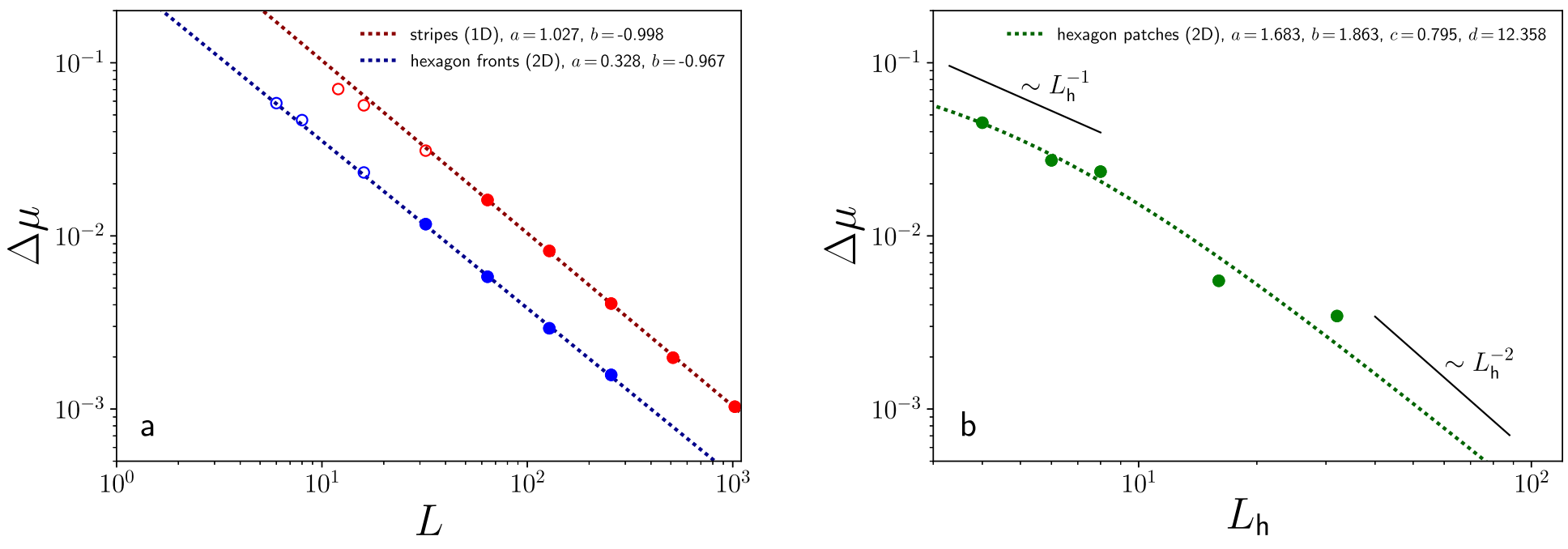

**Figure 14.** The width $\Delta\mu$ of the zigzag line for (a) localized 1D stripes, localized 2D hexagonal fronts and (b) localized 2D hexagonal patches as a function of the domain size ($L$ and $L_h$, respectively) with line styles as given in the legend. The circles indicate the $\Delta\mu$ values extracted from the zigzag lines in figures 10, 13 and 16 while the dashed lines show (a) the power law fits $\Delta\mu(L) = aL^b$ and (b) the fit $\Delta\mu(L_h) = a(L_h^b + 4L_h^c + d)^{-1}$ with the parameters $a, b, c, d$ provided in the legends. Only the filled circles are used to generate the fits.

To strengthen our argument that the different scaling behaviors can be attributed to the fully 2D nature of the hexagonal patches, we consider stripe-like arrangements of hexagonally ordered bumps. See [64, 65] for an analysis of similar states in the case of the standard nonconserved Swift–Hohenberg equation. We find that these exhibit scaling behavior that is identical to that of truly 1D structures. Figures 15 and 16 show the corresponding $(\phi_0, \mu)$ bifurcation diagrams, the associated zigzag curves and the scaling of $\Delta\mu$. These plots are computed for domains of size $L_x \times L_y$ with $L_y = L_y^c = 4\pi/\sqrt{3}$ and different values of $L_x$. Note that as in the 1D case there are two branches of localized states: one with an odd number of rows of bumps (green lines in figures 15(a) and (b), profiles in figure 15(c)) and one with an even number (red lines in figures 15(a) and (b), profiles in figure 15(d)).

These results show that quasi-1D localized structures behave in a manner that is very similar to strictly 1D structures; the approach to the Maxwell construction in the limit $L_x \to \infty$ is therefore also similar although, naturally, the coexistence value of the chemical potential that is approached corresponds to the 2D Maxwell line for hexagonal crystal-liquid coexistence and not the 1D Maxwell line for stripe-liquid coexistence. In particular, we expect that in this case the resulting Maxwell curve consists of identical linear segments of vanishing length and is likewise nowhere differentiable. However, the case of fully 2D hexagonal structures is different (figure 13), since there the zigzag segments have different lengths, although all appear once again to have the same slope. This distribution of lengths reflects the different effects on the chemical potential of adding bumps in the middle of a face (if the face has an even number of bumps) or two bumps on either side of the middle (if it has an odd number of bumps) and then of adding further bumps at the remaining sites along each face. These differences from the quasi-1D case imply that the zigzag structure does not repeat as one traverses the zigzag curve from one end to the other, a fact that is responsible for the departure of the exponent $b$ from its 1D value $b = -1$.

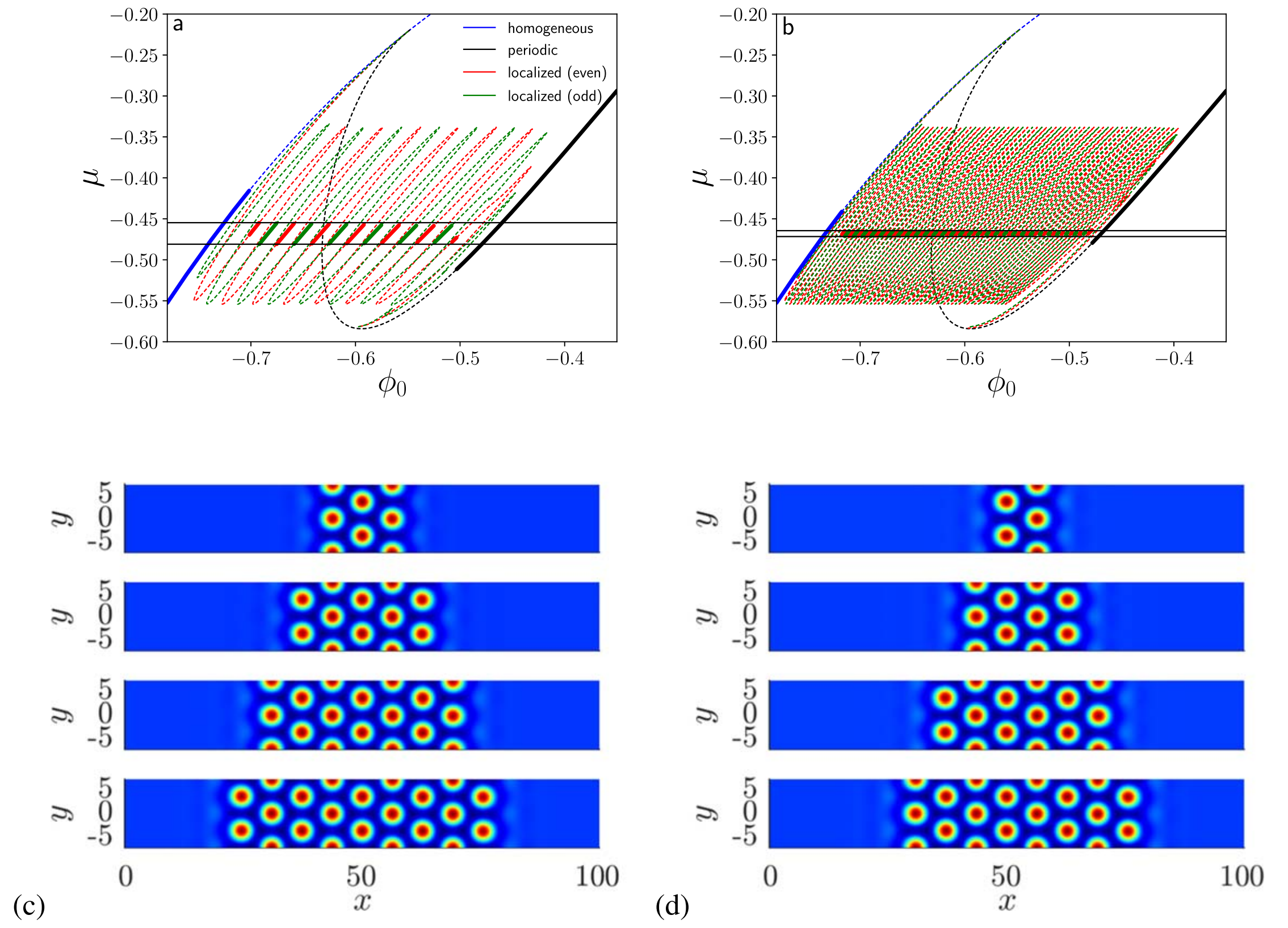

**Figure 15.** (a), (b) Bifurcation diagrams for 2D hexagonal fronts in a domain of size $L_x \times L_y$ showing the chemical potential $\mu$ as a function of the mean concentration $\phi_0$ for $L_y = L_y^c$ and (a) $L_x = 8L_x^c$, (b) $L_x = 32L_x^c$ with $L_x^c = 4\pi$. Shown are snaking branches of both odd and even hexagonal fronts as well as the uniform state and the hexagonal crystal. The horizontal lines indicate the interval $\Delta\mu$ in $\mu$ where the various localized states correspond to global energy minima: $\Delta\mu = 0.023\,252$ in (a) and $\Delta\mu = 0.005\,812$ in (b). The four subpanels in (c) and (d) show selected profiles of the 2D hexagonal front solutions when $L_x = 8L_x^c$ along the snaking branch with an odd (green branch in (a), (b)) and an even (red branch in (a), (b)) number of rows, respectively. To provide a clearer indication of the structure we show the profiles on an $L_y = 2L_y^c$ domain. Subpanels in (c) correspond to, from top to bottom, $\mu = -0.5057$, $\phi_0 = -0.7077$; $\mu = -0.5002$, $\phi_0 = -0.6718$; $\mu = -0.4766$, $\phi_0 = -0.6226$ and $\mu = -0.4782$, $\phi_0 = -0.5905$, while those in (d) correspond to $\mu = -0.4478$, $\phi_0 = -0.6877$; $\mu = -0.4631$, $\phi_0 = -0.6637$; $\mu = -0.4430$, $\phi_0 = -0.6160$ and $\mu = -0.4678$, $\phi_0 = -0.5996$.

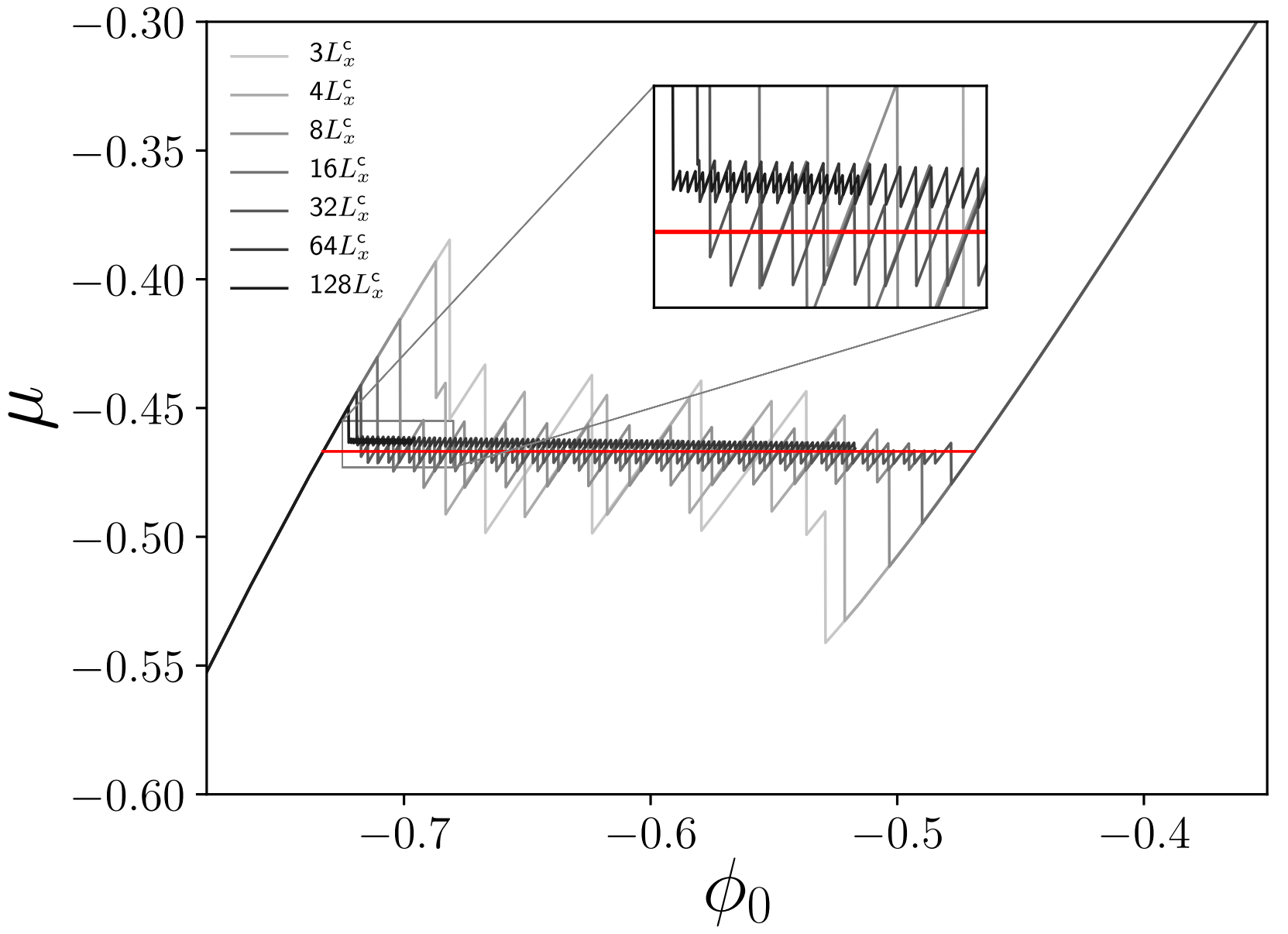

**Figure 16.** As in figures 15(a), (b) but showing the zigzag curves consisting of global minima of the free energy density only, for various strip lengths $L_x$ from $3L_x^c$ to $128L_x^c$ as given in the legend. The horizontal red line indicates the Maxwell construction as obtained in section 4.1. The slight offset between this and the zigzag curves for larger strip lengths is due to unavoidable numerical discretization errors.

Note, finally, that the inset in figure 16 reveals that the zigzag curves for $64L_x^c$ and $128L_x^c$ are both slightly off the exact coexistence value of $\mu$ indicated by the red horizontal line. We believe that this results from the resolution of the spatial grid that cannot be sufficiently refined for full convergence. This effect is also visible in the curve for $L_h = 16L_c^h$ in figure 13. Despite this, $\Delta\mu$ still follows a clear power law (figure 14(a)).

## 5. Conclusions

In this paper we have revisited the Maxwell construction that predicts the location of a first order phase transition in thermodynamic systems at finite temperature. Our work sheds new light on the process whereby a finite size system approaches the TL.

We have considered two basic mean field models, a phase separation model described by the Cahn–Hilliard equation and the PFC model or conserved Swift–Hohenberg equation that describes the process of crystallization from a melt. The former case is simpler since the two phases involved are both uniform. As a result any departure from the TL arises from the presence of interfaces whose contribution to the free energy vanishes in this limit. Despite this well-known property our analysis sheds new light on the TL: for example, in 2D it shows that the Maxwell construction (which we recover) involves different states (stripe and cluster states) depending on the mean concentration $\phi_0$ (see figure 6). In particular, in finite size domains, however large, the minimum energy state does not have the same spatial structure for all values of $\phi_0$.

The second model, the PFC model, reveals additional complexity. In 1D, finite systems are characterized by a large number of spatially localized states lying on a pair of intertwined branches that straddle the Maxwell point between the homogeneous (liquid) phase and the periodic (crystalline) phase. As in nonconserved systems these states gain and lose linear stability at successive folds along these snaking branches, and in a $\phi_0(\mu)$ plot these folds are located at particular values $\mu^\pm$ of the chemical potential ($\mu^- < \mu_{\mathrm{Maxwell}} < \mu^+$), i.e. the folds are aligned [20]. This behavior is independent of the domain length $L$ once $L$ is large enough, provided the localized structures remain localized, and do not fill the whole domain (figure 9). However, what does depend on $L$ is the number of available states which increases in proportion to $L$ generating more and more oscillations across the Maxwell point. In addition, the interval in $\mu$, $\Delta\mu$, within which these states correspond to the global free energy minimum decreases to zero, to good accuracy, as $L^{-1}$. This is because the minimum energy state jumps discontinuously to longer and longer localized structures as $\phi_0$ increases and each of these states remains a global minimum for the same $\mu$ interval $\Delta\mu$. In the limit $L \to \infty$ one therefore finds that the Maxwell line consists of a dense set of transitions between different and successively longer localized states. Thus in this case, too, the Maxwell line corresponds to a succession of distinct states (figure 10).

In 2D the PFC model is even more interesting since the cluster states now typically consist of hexagonal patches. In a $\phi_0(\mu)$ plot these patches snake in the same way as in nonconserved systems [20], implying that they grow by adding a bump in the middle of each interface followed by the addition of bumps on either side of these first bumps until a new and larger patch has been created. Because of this gradual increase in area the growth process does not repeat in a periodic manner, with the number of back and forth oscillations across the Maxwell line increasing with $\phi_0$. Here too the interval $\Delta\mu$ within which each patch corresponds to minimum free energy is well defined and it too decreases to zero as the domain area grows. For hexagonal domains we find that $\Delta\mu \sim L_h^{-2}$, where $L_h$ is the domain scale (figure 14). As in the 1D case in the limit $L \to \infty$ we recover the Maxwell construction, with the Maxwell line consisting of a dense set of transitions between distinct hexagonal patches (figure 13). We have argued that the different scaling exponent, $b \sim -2$, is a consequence of the 2D nature of the patches since a similar analysis of hexagonal stripes, i.e. localized regions of hexagons with only one extended direction, exhibit the same $\Delta\mu$ scaling as the strictly 1D problem.

Both of the mean field models we have considered can be derived as local (gradient expansion) approximations of density functional theory (DFT) [19, 38, 66, 67]. Formally, DFT is derived by averaging over all states of the system, i.e. including all thermal fluctuation effects. However, since in practice one must make an approximation for the free energy functional, the mean field models do not include the full contribution of thermal fluctuations. However, since at least some fluctuation contributions are already included and averaged over, one should not add additional fluctuating terms. Instead, to obtain a quantitatively more accurate theory, one should employ better approximations for the free energy functional. This would result in quantitatively different results, but the qualitative behavior regarding the emergence of the Maxwell construction would likely remain unchanged. With these mean field models complete bifurcation diagrams can be determined including the unstable and metastable states that computer simulations are normally unable to capture. To see how the full spectrum of fluctuations amends the picture one would need to pair our approach with large scale computer simulations, as done in the context of partially wetting droplets on solid substrates in [68].

Finally, we discuss another limitation of our study. We have considered finite size systems with ideal boundaries, i.e. periodic or Neumann boundaries (for an extensive discussion of their relation see [69]). The latter do not influence the phase behavior at the boundaries, but only break the translation invariance of an infinite system. An interesting question for future work concerns the role played by rigid boundaries in the approach to the TL. In the case of mixtures these may represent preferential adsorption of one component or a changed interaction between components at a boundary [70]. Such boundaries may result in a very rich surface phase behavior (see e.g. figure 6 in [71] for a phase diagram, figures 3–14 in [53] for bifurcation diagrams and 1D concentration profiles, and [58] for 2D results). Incorporating such boundaries into the present study will most likely add a new level of complexity to some aspects of the bifurcation diagrams, allowing one to investigate the interplay between different bulk (liquid-gas, demixing, crystallization) and surface (wetting, pre-wetting, surface freezing, pre-melting) phase transitions in finite systems and the corresponding transition towards the TL.

## Acknowledgments

UT thanks the Physics Department at the University of California at Berkeley and the Mathematics Department at the University of British Columbia for their hospitality during the finalization of the manuscript. The work of EK was supported in part by the National Science Foundation (USA) under grants DMS-1613132 and DMS-1908891 while the work of AJA was supported by the EPSRC under grant EP/P015689/1. We acknowledge support by the Open Access Publication Fund of the University of Münster.

## ORCID iDs

Uwe Thiele https://orcid.org/0000-0001-7989-9271
Andrew J Archer https://orcid.org/0000-0002-4706-2204

## References

[1] Kondepudi D and Prigogine I 1998 *Modern Thermodynamics—From Heat Engines to Dissipative Systems* (Chichester: Wiley)
[2] Stanley H E 1987 *Introduction to Phase Transitions and Critical Phenomena* (Oxford: Oxford University Press)
[3] Nicolis G 1999 *Introduction to Nonlinear Science* (Cambridge: Cambridge University Press)
[4] Pismen L M 2006 *Patterns and Interfaces in Dissipative Dynamics* (Berlin: Springer)
[5] Strogatz S H 2014 *Nonlinear Dynamics and Chaos* (New York: Westview Press)
[6] Meakin P 1998 *Fractals, Scaling and Growth far from Equilibrium* (*Cambridge Nonlinear Science Series* vol 5) (Cambridge: Cambridge University Press)
[7] Budd C J, Hunt G W and Kuske R 2001 Asymptotics of cellular buckling close to the Maxwell load *Proc. R. Soc.* A **457** 2935–64
[8] Burke J and Knobloch E 2006 Localized states in the generalized Swift–Hohenberg equation *Phys. Rev.* E **73** 056211
[9] Burke J and Knobloch E 2007 Homoclinic snaking: structure and stability *Chaos* **17** 037102
[10] Coullet P 2002 Localized patterns and fronts in nonequilibrium systems *Int. J. Bifurcation Chaos* **12** 2445–57
[11] Lloyd D J B, Gollwitzer C, Rehberg I and Richter R 2015 Homoclinic snaking near the surface instability of a polarisable fluid *J. Fluid Mech.* **783** 283–305
[12] Hoyle R B 2006 *Pattern Formation—An Introduction to Methods* (Cambridge: University Press)
[13] Knobloch E 2015 Spatial localization in dissipative systems *Annu. Rev. Condens. Matter Phys.* **6** 325–59
[14] Hunt G W, Peletier M A, Champneys A R, Woods P D, Ahmer Wadee M, Budd C J and Lord G J 2000 Cellular buckling in long structures *Nonlinear Dyn.* **21** 3–29
[15] Bray A J 1994 Theory of phase-ordering kinetics *Adv. Phys.* **43** 357–459
[16] Cahn J W 1965 Phase separation by spinodal decomposition in isotropic systems *J. Chem. Phys.* **42** 93–9
[17] Cahn J W and Hilliard J E 1958 Free energy of a nonuniform system: I. Interfacial free energy *J. Chem. Phys.* **28** 258–67
[18] Elder K R and Grant M 2004 Modeling elastic and plastic deformations in nonequilibrium processing using phase field crystals *Phys. Rev.* E **70** 051605
[19] Emmerich H, Löwen H, Wittkowski R, Gruhn T, Tóth G I, Tegze G and Gránásy L 2012 Phase-field-crystal models for condensed matter dynamics on atomic length and diffusive time scales: an overview *Adv. Phys.* **61** 665–743
[20] Thiele U, Archer A J, Robbins M J, Gomez H and Knobloch E 2013 Localized states in the conserved Swift–Hohenberg equation with cubic nonlinearity *Phys. Rev.* E **87** 042915
[21] Bernard E P and Krauth W 2011 Two-step melting in two dimensions: first-order liquid-hexatic transition *Phys. Rev. Lett.* **107** 155704
[22] Binder K, Block B J, Virnau P and Troster A 2012 Beyond the van der Waals loop: what can be learned from simulating Lennard-Jones fluids inside the region of phase coexistence *Am. J. Phys.* **80** 1099–109
[23] MacDowell L G, Shen V K and Errington J R 2006 Nucleation and cavitation of spherical, cylindrical, and slablike droplets and bubbles in small systems *J. Chem. Phys.* **125** 034705
[24] MacDowell L G, Virnau P, Müller M and Binder K 2004 The evaporation/condensation transition of liquid droplets *J. Chem. Phys.* **120** 5293

[25] MacDowell L G 2011 Computer simulation of interface potentials: towards a first principle description of complex interfaces *Eur. Phys. J. Spec. Top.* **197** 131
[26] Engelnkemper S, Gurevich S V, Uecker H, Wetzel D and Thiele U 2019 Continuation for thin film hydrodynamics and related scalar problems *Computational Modeling of Bifurcations and Instabilities in Fluid Mechanics* (*Computational Methods in Applied Sciences*) ed A Gelfgat vol 50 (Berlin: Springer) pp 459–501
[27] Pattamatta S, Elliott R S and Tadmor E B 2014 Mapping the stochastic response of nanostructures *Proc. Natl Acad. Sci. USA* **111** E1678–86
[28] Dupuy L M, Tadmor E B, Miller R E and Phillips R 2005 Finite-temperature quasicontinuum: molecular dynamics without all the atoms *Phys. Rev. Lett.* **95** 060202
[29] Doi M 2013 *Soft Matter Physics* (Oxford: Oxford University Press)
[30] Thiele U 2018 Recent advances in and future challenges for mesoscopic hydrodynamic modelling of complex wetting *Colloids Surf.* A **553** 487–95
[31] Archer A J and Evans R 2004 Dynamical density functional theory and its application to spinodal decomposition *J. Chem. Phys.* **121** 4246–54
[32] Langer J S 1992 An introduction to the kinetics of first-order phase transitions *Solids Far from Equilibrium* ed C Godrèche (Cambridge: Cambridge University Press) pp 297–363
[33] Chaikin P M and Lubensky T C 1995 *Principles of Condensed Matter Physics* (Cambridge: Cambridge University Press)
[34] Foard E M and Wagner A J 2012 Survey of morphologies formed in the wake of an enslaved phase-separation front in two dimensions *Phys. Rev.* E **85** 011501
[35] Goh R and Scheel A 2015 Hopf bifurcation from fronts in the Cahn–Hilliard equation *Arch. Ration. Mech. Anal.* **217** 1219–63
[36] Köpf M H, Gurevich S V, Friedrich R and Thiele U 2012 Substrate-mediated pattern formation in monolayer transfer: a reduced model *New J. Phys.* **14** 023016
[37] Elder K R, Katakowski M, Haataja M and Grant M 2002 Modeling elasticity in crystal growth *Phys. Rev. Lett.* **88** 245701
[38] Archer A J, Ratliff D J, Rucklidge A M and Subramanian P 2019 Deriving phase field crystal theory from dynamical density functional theory: consequences of the approximations *Phys. Rev.* E **100** 022140
[39] Archer A J, Robbins M J, Thiele U and Knobloch E 2012 Solidification fronts in supercooled liquids: how rapid fronts can lead to disordered glassy solids *Phys. Rev.* E **86** 031603
[40] van Teeffelen S, Backofen R, Voigt A and Löwen H 2009 Derivation of the phase-field-crystal model for colloidal solidification *Phys. Rev.* E **79** 051404
[41] Greenwood M, Rottler J and Provatas N 2011 Phase-field-crystal methodology for modeling of structural transformations *Phys. Rev.* E **83** 031601
[42] Wu K A, Plapp M and Voorhees P W 2010 Controlling crystal symmetries in phase-field crystal models *J. Phys.: Condens. Matter* **22** 364102
[43] Cross M C and Hohenberg P C 1993 Pattern formation out of equilibrium *Rev. Mod. Phys.* **65** 851–1112
[44] Dijkstra H A *et al* 2014 Numerical bifurcation methods and their application to fluid dynamics: analysis beyond simulation *Commun. Comput. Phys.* **15** 1–45
[45] Doedel E, Keller H B and Kernevez J P 1991 Numerical analysis and control of bifurcation problems (I): bifurcation in finite dimensions *Int. J. Bifurcation Chaos* **1** 493–520
[46] Doedel E J and Oldeman B E 2009 *AUTO07p: Continuation and Bifurcation Software for Ordinary Differential equations* (Montreal: Concordia University)
[47] Thiele U, Kamps O and Gurevich S V (ed) 2014 *Münsteranian Tutorials on Nonlinear Science: Continuation* (Münster: CeNoS) (http://uni-muenster.de/CeNoS/Lehre/Tutorials)
[48] Uecker H and Wetzel D 2014 Numerical results for snaking of patterns over patterns in some 2d Selkov–Schnakenberg reaction-diffusion systems *SIAM J. Appl. Dyn. Syst.* **13** 94–128
[49] Uecker H, Wetzel D and Rademacher J 2014 pde2path—a Matlab package for continuation and bifurcation in 2D elliptic systems *Numer. Math.-Theory Methods Appl.* **7** 58–106
[50] Thiele U and Knobloch E 2004 Thin liquid films on a slightly inclined heated plate *Physica* D **190** 213–48
[51] Thiele U, Vega J M and Knobloch E 2006 Long-wave Marangoni instability with vibration *J. Fluid Mech.* **546** 61–87
[52] Novick-Cohen A and Segel L A 1984 Nonlinear aspects of the Cahn–Hilliard equation *Physica* D **10** 277–98
[53] Thiele U, Madruga S and Frastia L 2007 Decomposition driven interface evolution for layers of binary mixtures: I. Model derivation and stratified base states *Phys. Fluids* **19** 122106
[54] Diez J A and Kondic L 2007 On the breakup of fluid films of finite and infinite extent *Phys. Fluids* **19** 072107
[55] Thiele U, Brusch L, Bestehorn M and Bär M 2003 Modelling thin-film dewetting on structured substrates and templates: bifurcation analysis and numerical simulations *Eur. Phys. J.* E **11** 255–71
[56] Gameiro M and Lessard J-P 2010 Analytic estimates and rigorous continuation for equilibria of higher-dimensional PDEs *J. Differ. Equ.* **249** 2237–68
[57] Maier-Paape S, Mischaikow K and Wanner T 2007 Structure of the attractor of the Cahn–Hilliard equation on a square *Int. J. Bifurcation Chaos* **17** 1221–63
[58] Bribesh F A M, Frastia L and Thiele U 2012 Decomposition driven interface evolution for layers of binary mixtures: III. Two-dimensional steady film states *Phys. Fluids* **24** 062109
[59] Brazovskii S A 1975 Phase transition of an isotropic system to a nonuniform state *J. Exp. Theor. Exp.* **41** 85
[60] Robbins M J, Archer A J, Thiele U and Knobloch E 2012 Modelling fluids and crystals using a two-component modified phase field crystal model *Phys. Rev.* E **85** 061408
[61] Dawes J H P 2008 Localized pattern formation with a large-scale mode: slanted snaking *SIAM J. Appl. Dyn. Syst.* **7** 186–206
[62] Pradenas B, Araya I, Clerc M G, Falcón C, Gandhi P and Knobloch E 2017 Slanted snaking of localized Faraday waves *Phys. Rev. Fluids* **2** 064401
[63] Knobloch E 2016 Localized structures and front propagation in systems with a conservation law *IMA J. Appl. Math.* **81** 457–87
[64] Lloyd D J B, Sandstede B, Avitabile D and Champneys A R 2008 Localized hexagon patterns of the planar Swift–Hohenberg equation *SIAM J. Appl. Dyn. Syst.* **7** 1049–100
[65] Avitabile D, Lloyd D J B, Burke J, Knobloch E and Sandstede B 2010 To snake or not to snake in the planar Swift–Hohenberg equation *SIAM J. Appl. Dyn. Syst.* **9** 704–33
[66] Evans R 1992 Density functionals in the theory of nonuniform fluids *Fundamentals of Inhomogeneous Fluids* ed D Henderson (New York: Marcel Dekker) pp 85–176

[67] Hansen J-P and McDonald I R 2013 *Theory of Simple Liquids: with Applications to Soft Matter* (New York: Academic)
[68] Tretyakov N, Müller M, Todorova D and Thiele U 2013 Parameter passing between molecular dynamics and continuum models for droplets on solid substrates: The static case *J. Chem. Phys.* **138** 064905
[69] Crawford J D, Golubitsky M, Gomes M G M, Knobloch E and Stewart I M 1991 Boundary conditions as symmetry constraints *Singularity Theory and its Applications, Part II* (*Lecture Notes in Mathematics* vol 1463) (New York: Springer) pp 63–79
[70] Fischer H P, Maass P and Dieterich W 1997 Novel surface modes in spinodal decomposition *Phys. Rev. Lett.* **79** 893–6
[71] Cahn J W 1977 Critical-point wetting *J. Chem. Phys.* **66** 3667–72